\documentclass[12pt,letterpaper]{article}%
\usepackage{amsmath}
\usepackage{amsfonts}
\usepackage{amssymb}
\usepackage{nopageno}
\usepackage{graphicx}%
\newtheorem{theorem}{Theorem}

\newtheorem{claim}[theorem]{Claim}

\newtheorem{conjecture}[theorem]{Conjecture}

\newtheorem{remark}[theorem]{Remark}

\makeatletter
\def\@oddhead{
\vbox{
\hbox to\hsize{\oddmarkA \oddmarkB \hfill \oddmarkC}
}
}
\makeatother
\def\oddmarkA{{\bf }{}}
\def\oddmarkB{}
\def\oddmarkC{\thepage}
\begin{document}

\title{\textbf{Part \ I. \ The Cosmological Vacuum }\\\textbf{from a Topological Perspective}}
\author{\textbf{R. M. Kiehn\thanks{http://www.cartan.pair.com}}\\Physics Department, University of Houston\\Retired emeritus to Mazan, France}
\maketitle

\begin{abstract}
\ This article examines how the physical presence of field energy and
particulate matter can be interpreted in terms of the topological properties
of space-time. The theory is developed in terms of vector and matrix equations
of exterior differential systems, which are not constrained by tensor
diffeomorphic equivalences. \ The first postulate defines the field properties
(a vector space continuum) of the Cosmological Vacuum in terms of matrices of
basis functions that map exact differentials into neighborhoods of exterior
differential 1-forms (potentials). \ The second postulate requires that the
field equations must satisfy the First Law of Thermodynamics dynamically
created in terms of the Lie differential with respect to a process direction
field acting on the exterior differential forms that encode the thermodynamic
system. The vector space of infinitesimals need not be global and its
compliment is used to define particle properties as topological defects
embedded in the field vector space. The potentials, as exterior differential
1-forms, are not (necessarily) uniquely integrable: the fibers can be twisted,
leading to possible Chiral matrix arrays of certain 3-forms defined as
Topological Torsion and Topological Spin. \ A significant result demonstrates
how the coefficients of Affine Torsion are related to the concept of Field
excitations (mass and charge); another demonstrates how thermodynamic
evolution can describe the emergence of topological defects in the physical vacuum.

\end{abstract}
\tableofcontents

\section{Preface}

In 1993-1998, Gennady Shipov\ \cite{Shipov} presented his pioneering concept
of a \textit{modified} A$_{n}$ space of "Absolute Parallelism". \ Shipov
implied that such geometric structures could exhibit "Affine Torsion",
$\left[  \mathbb{C(}x^{a})\right]  \symbol{94}\left\vert dx^{k}\right\rangle
$, where $\left[  \mathbb{C(}x^{a})\right]  $ is the Cartan matrix of
Connection 1-forms. \ Such\ non-Riemannian structures are different from those
induced by Gauss curvature effects associated with Riemannian gravitational
fields, for the A$_{n}$ space defines a domain of zero Gauss curvature.
\ Although others had considered spaces with torsion, Shipov's work\ was
important to me for it stimulated the realization that the \textit{physical}
vacuum is not a void of nothingness, but indeed is something that can have
structure and field properties that are physically measurable \cite{rmkpv}.
\ An objective of this article is to demonstrate how "Affine Torsion", defined
in terms of Cartan's matrix of connection 1-forms, $\left[  \mathbb{C(}%
x^{a})\right]  \symbol{94}\left\vert dx^{k}\right\rangle $, leads to
interesting thermodynamic and topological conclusions. \ From a topological
perspective, the similarities between the vector field continuum properties of
hydrodynamics and electrodynamics are remarkable. \ The continuum fields with
pair, or impair, Affine Torsion 2-forms can delineate topologically between
the positive definite property of mass and the indefinite property of charge.
\ In addition, pair and impair 3-forms of Topological Torsion and Topological
Spin are to be found in non-equilibrium hydrodynamic systems as well as in
non-equilibrium electromagnetic systems. \ The combination of the two
continuum fields leads to the Cosmological Vacuum.

Recall that Arnold Sommerfeld in his formulation \cite{Sommerfeld}\ of
electromagnetism described quantities of field excitations (think D and H
related to sources), now known to be representable as coefficients of certain
exterior 2-form (impair) densities, $G$. \ \ In general, $dG$ is not zero and
produces a 3-form (impair) density of "charge-current", $J=dG$. \ On closed
domains where $dG=J=0$, closed integrals\ of $G$ over cycles (not a
boundary)$\ $define the number, $n$, of unit charges, $e$, contained within
the closed cycle: $\int\int_{2dcycle}G=ne$. \ Integrals over a boundary in the
closed domains are zero, which implies that the charges confined to different
cycles can be of opposite sign, and add to zero. \ Charge, although
topologically quantized, has both plus and minus values. \ For every positive
quantity of charge there is an equal and opposite negative quantity\ of
charge, in agreement with the physical conclusion that the electrodynamic
universe is charge neutral \ in an isolated-equilibrium configuration. \ This
result is due to the fact that impair densities are sensitive to the sign of
the determinant of a 4-volume transformation; that is, impair densities are
sensitive to the choice of orientation. \ 

Sommerfeld also described fields of intensity, now known to be coefficients of
(pair) 2-forms, $F$ (think E and B\ \symbol{126} related to forces). \ The
2-forms, $F$, are exact, and therefor are closed over the domain of
C2\ definition. \ In thermodynamic terms, the additive objects of quantity,
$G$, are homogeneous of degree one, and the objects of intensity, $F$, are
homogeneous of degree zero. \ In deRham's terminology, electromagnetic
excitations, $G$, are "impair" (exterior differential form densities whose
closed integrals are sensitive to orientation) and the intensities, $F$, are
"pair" (exterior differential forms whose closed integrals do not depend upon
orientation). \ It is the "impair" feature that leads to charges of different
sign. \ These developments in topological electrodynamics (see Vol. 4
\cite{rmklulu}) guide the development for topological hydrodynamics.
\ However, there are some important differences. \ 

Historically, charge, like mass, had been presumed to be a scalar, and
therefore should not be orientation dependent. Closed impair densities, $G$,
yield a pseudoscalar integral $\int\int_{2dcycle}G$ when integrated over a
closed, not bounded, domain. \ The values of the closed integrals have a sign
dependent upon the $\pm$ choice of orientation \cite{PostQR}.\ \ The
historical assumptions of charge as a scalar are not compatible with the
topological format that $G$ is impair. \ E. Post, through his studies of
magnetic effects in crystals, demonstrated long ago that if charge was a
scalar, magnetic permeability would vanish in crystals that had a center of
symmetry, counter to experimental observation of such effects\ \cite{PostTG}.
\ Post, over the years has repeatedly championed this fact that charge is a
pseudoscalar, but only recently has the physical community started to take
note of this important, experimentally confirmed, result \cite{HEHLF}. \ Most
of the physics community still hangs on to the dogma that charge is a scalar. \ 

As mass is considered to be a positive definite quantity, it should not be
presented as having period integrals of 2-forms that are impair. \ Instead the
mass-current 3-forms must be defined as pair 3-form densities, which are not
sensitive to orientation, a fact that distinguishes them from the
charge-current impair 3-form densities. \ What is the reason for such
differences? \ Mirror images of mass are of the same sign.

One of the most significant results of the current work herein, which goes
beyond Shipov's focus on possible physical properties of the "vacuum", is the
fact that the vector of Affine Torsion 2-forms, $\left[  \mathbb{C(}%
x^{a})\right]  \symbol{94}\left\vert dx^{k}\right\rangle ,$ can consist of
either pair 2-form densities, $\left\vert \mathfrak{T}\right\rangle $, or
impair 2-form densities, $\left\vert G\right\rangle .$

\ Many investigations that appear in the literature force a self-duality (or
even anti-self-duality) on the distinct thermodynamic concepts of quantities
and intensities; these constraints can limit the thermodynamic generality and
applicability of the constrained theories. \ Indeed, in classic hydrodynamics,
the analog of the Affine Torsion 2-forms are a missing link, for the
theoretical assumptions (based on classical elasticity theory) which invoke
symmetric metric hypotheses (the definition of stress) do not permit the
generation of Affine Torsion in classical fluids. \ Once it is recognized that
the continuum fields of the physical vacuum need not be a vector space
generated by "symmetric" collineations, then fluids, as well as plasmas, can
support Affine Torsion or its equivalent: \ to repeat, the topological
theories of both hydrodynamics and electrodynamics, formally, are almost the
same! \ In much of this article, both the hydrodynamic notation and the
electrodynamic notation will be used side by side. \ For most readers I expect
the electrodynamic notation (in its engineering format of $\mathbf{E}$ and
$\mathbf{B}$ fields) will be the most readily comprehended. \ Over the years
the concepts of fluids with Affine Torsion structure have stimulated interest
in General Relativity, in superconductivity, and even in string theory: \ the
current buzz-word is: "spin fluids" \cite{Smalley}, \cite{Graf}. \ 

In electromagnetism the idea of "string-like" particles was utilized, if not
invented, by Bateman (1913): "...According to this idea a corpuscle has a kind
of tube or a thread attached to it...." (see p131, \cite{Bateman}). \ What
Bateman recognized was that there were two classes of solutions (representing
corpuscles of two types !) of the fundamental equations of electromagnetism.
\ The first solution class involves the classic vector ideas, and is related
to the standard constitutive ideas relating the two thermodynamic species:
$\mathbf{D}\ =\varepsilon\mathbf{E}$ and $\mathbf{B}\ =\mu\mathbf{H}$. \ The
second solution class was constructed in terms of complex solutions with
"zero" length. \ It is remarkable that this second model of a corpuscle could
be related to a "self dual" constitutive relationship of the form,
$\mathbf{D}=-\sqrt{-1}\gamma\mathbf{B}$ and $\mathbf{H}=\sqrt{-1}%
\gamma\mathbf{E}$. \ At the time, Bateman did not recognized that the second
solutions were what Cartan (a bit later) defined as Isotropic Spinors.
\ Bateman also failed to realize that such Spinor solutions were generators of
conjugate pairs of minimal surfaces (tangential discontinuities). \ In
addition, it is known that minimal surfaces in a domain with a Minkowski
signature can exhibit structures that appear to be branes connected by strings
(see Vol 2. \cite{rmklulu}).

\begin{center}%
%TCIMACRO{\FRAME{itbpF}{3.5812in}{2.8314in}{0in}{}{}{branemay7.wmf}%
%maintain-aspect-ratio TRUE;  display "PICT";  valid_file "F";
%width 3.5812in;  height 2.8314in;  depth 0in;  original-width 3.9479in;
%original-height 3.1142in;  cropleft "0";  croptop "1";  cropright "1";
%cropbottom "0";  filename 'branemay7.wmf';file-properties "XNPEU";}}}%
%BeginExpansion
{\includegraphics[
height=2.8314in,
width=3.5812in
]%
{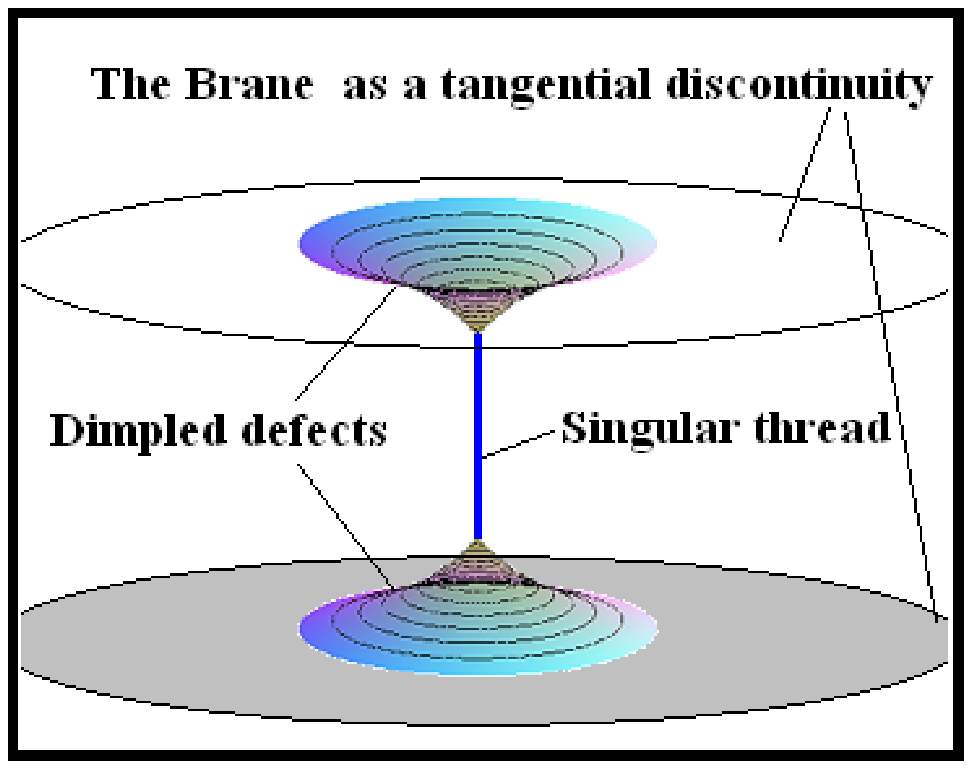}%
}%
%EndExpansion
\ 

\textbf{Figure 1. \ Minimal surfaces (tangential discontinuities}s)

in a space with a Minkowski signature.
\end{center}

Similar string like ideas were developed by Post who assumed that there were
paths of connected points that could represent "collectively" the electron,
and that these "topologically collective paths" could be described as "cyclic
time" orbits, when evaluated in terms of closed integrals of closed 2-forms.
\ The classic solution to the Gauss period integral corresponds to charges
within volumes contained by 2D--cycles. \ There are however other solutions
that correspond to linking of pairs, or knotting of single cyclic paths, and
even to "points". \ These ideas can be extended to spaces of Pfaff Topological
Dimension 3, with both impair and pair 3-dimensional period integrals, leading
to both the linking of charge triplets of cyclic time paths and the linking of
mass triplets of cyclic time paths. \ 

String theorists in elementary particle physics speculate about brane walls
being connected by "strings", and introduce concepts such as "dilatons" and
"axions"\footnote{For dilatons, think expansion about a fixed point. \ For
axions, think rotations about a fixed point.}, without any experimental
evidence of their existence. \ All of this is done without regard to
macroscopic "engineering" examples of fluids that can support Affine Torsion,
but are fluids (or plasmas) in non-equilibrium configurations. \ Fluids that
admit Affine Torsion can produce macroscopic structures that match the
verbiage of elementary particle "string theorists". \ Perhaps the most vivid
(and most easily produced) macroscopic example of strings and branes is given
by the creation of Falaco Solitons (see Vol. 2 \cite{rmklulu}). \ Falaco
Solitons are minimal surface dimples (branes) connected by strings. \ Similar
structures were observed by Hopfinger \cite{Hopfinv} in a rotating tank of a
turbulent fluid, which produced Hasimoto vortex kinks. \ \ Models of the
photon can be constructed with similar topological configurations
\cite{rmkspie}. \ Indeed, the chemical make up of soap films appears to be
that of a double layer (branes) connected by molecular strings. \ These
examples are real world exhibitions of a string theory, but they do not
require many dimensions, or many worlds,\ or quantum mechanics, but they do
suggest that "string theorists"\ should apply their kraft to the real world.
\ In the examples that follow, the common connection between these string and
particle concepts relates to the ability to construct homogeneous exterior
differential forms which are closed.

In summary, this article examines how the physical presence of field energy
and particulate matter could influence the \textit{topological} properties
(not only the \textit{geometrical} properties) of space-time to form a
"Cosmological Vacuum". \ The field part of the Cosmological Vacuum is defined
as a non-global 4D vector space, a continuum, in which are embedded
topological defect structures that play the role of particles. \ It becomes
apparent that the topological method is not intrinsically dependent on size or
shape, and therefor can be used as a universal tool for studying
non-equilibrium continuum systems at all scales. \ 

\subsection{The Point of departure}

The point of departure in this article consists of three parts:

\textbf{I}. \ Shipov's constraint of Absolute Parallelism, is extended to
include a larger set of admissible systems. \ The larger set is based on the
sole requirement that \textit{infinitesimal} neighborhoods of a "Cosmological
Vacuum" are elements of a 4D vector space, defined by the mapping of a vector
arrays of exact differentials into a vector arrays of non-exact differential
1-forms. \ Some authors have called this a "local mapping", but in order to
preserve the topological implications, I prefer the words "infinitesimal
mappings". \ The method permits the inclusion of non-trivial bundles when the
1-forms are not integrable

String theorists in elementary particle physics speculate about brane walls
being connected by "strings", and introduce concepts such as "dilatons" and
"axions"\footnote{For dilatons, think expansion about a fixed point. \ For
axions, think rotations about a fixed point.}, without any experimental
evidence of their existence. \ All of this is done without regard to
macroscopic "engineering" examples of fluids that can support Affine Torsion,
but are fluids (or plasmas) in non-equilibrium configurations. \ Fluids that
admit Affine Torsion can produce macroscopic structures that match the
verbiage of elementary particle "string theorists". \ Perhaps the most vivid
(and most easily produced) macroscopic example of strings and branes is given
by the creation of Falaco Solitons (see Vol. 2 \cite{rmklulu}). \ Falaco
Solitons are minimal surface dimples (branes) connected by strings. \ Similar
structures were observed by Hopfinger \cite{Hopfinv} in a rotating tank of a
turbulent fluid, which produced Hasimoto vortex kinks. \ \ Models of the
photon can be constructed with similar topological configurations
\cite{rmkspie}. \ Indeed, the chemical make up of soap films appears to be
that of a double layer (branes) connected by molecular strings. \ These
examples are real world exhibitions of a string theory, but they do not
require many dimensions, or many worlds,\ or quantum mechanics, but they do
suggest that "string theorists"\ should apply their kraft to the real world.
\ In the examples that follow, the common connection between these string and
particle concepts relates to the ability to construct homogeneous exterior
differential forms which are closed.

In summary, this article examines how the physical presence of field energy
and particulate matter could influence the \textit{topological} properties
(not only the \textit{geometrical} properties) of space-time to form a
"Cosmological Vacuum". \ The field part of the Cosmological Vacuum is defined
as a non-global 4D vector space, a continuum, in which are embedded
topological defect structures that play the role of particles. \ It becomes
apparent that the topological method is not intrinsically dependent on size or
shape, and therefor can be used as a universal tool for studying
non-equilibrium continuum systems at all scales. \ \ Sometimes the mapping is
referred to as the Lie algebra.

\ \ The arbitrary matrix of C2 functions with non-zero determinant (that
establishes the vector space mapping of differentials) is defined as a Basis
Frame, $\left[  \mathbb{B(}x^{a})\right]  $. \ The class of matrices with
non-zero determinant falls into two disjoint sets: \ those for which the
determinant is greater than zero, and those for which the determinant is less
than zero. \ The compliment of the vector space is then defined as those
subspace domains where the determinant of $\left[  \mathbb{B(}x^{a})\right]  $
is equal to zero. \ In other words, the vector space representing the
continuum field properties of the Cosmological Vacuum need not be global.
\ When a global object (like a Lie group) is replaced by its infinitesimal
version, (the Lie "infinitesimal group"), the system is called the Lie algebra.\ 

The basis vectors that make up the collineation Basis Frame for
\textit{infinitesimal} neighborhoods exhibit topological, differential
closure. \ That is, the differential of any column vector of the Basis Frame
is a linear combination of all of the column vectors that make up the Basis
Frame. \ The set of admissible Basis Frames for the vector space of
infinitesimal neighborhoods is richer than the set of basis frames for global
neighborhoods. \ The infinitesimal maps need not be integrable, and therefor
could represent non-trivial bundle concepts. \ In this article, the matrix
format of exterior differential forms is chosen as the mathematical vehicle of
choice, thus removing the "debauch des indices" associated with tensor
analysis, and admitting evolutionary processes that are not diffeomorphisms.
\ Moreover, the "twisting" of the "fibers" producing chiral effects (often
missed by classical tensor methods) becomes evident, as for any $\left[
\mathbb{B(}x^{a})\right]  $ there are usually two connections that satisfy the
conditions of differential closure. \ \qquad

The Basis Frames of interest are not necessarily Symmetric or Orthonormal,
which are properties associated with specific gauge constraints imposed on the
general system of infinitesimal mappings. \ This topological point of view
emphasizes the concept of a connection, and minimizes investigation of the
concepts that depend upon a metric. \ However it is recognized that the
concept of a signature is of importance to thermodynamic evolution, due to the
production of conjugate spinors. \ The orthogonal group, O(2n), associated
with a signature (+ + + +), preserves the euclidean structure. \ The General
Linear group, associated with a signature (+++ -), of complex elements
preserves the complex structure \cite{Arnold}. \ The Symplectic group can be
associated with the signature (- - - +). \ It is to be recognized that the
signature (+++ -) leads to Majorana Spinors, and the signature (- - - +) leads
to Dirac Spinors. \ 

\textbf{II}. \ In certain domains of base variables $\{x\}$ the Basis Frame
matrix, $[\mathbb{B}(x)],$ can be singular, and then one or more of its
four\ (possibly complex) eigenvalues is zero. \ These singular domains (or
objects) may be viewed as topological defects of 3 (topological) dimensions -
or less - embedded in the field domain of a 4 dimensional "Cosmological
Vacuum". \ These topological defects can be thought of as condensates, or
particles, or field discontinuities, or Spinor Null spaces. \ The major theme
of this article examines the continuum field properties of the "Cosmological
Vacuum", which is the domain free of singularities of the type\ $\det
[\mathbb{B}]=0$. \ The Basis Frame Matrix $[\mathbb{B}]$ will be assumed to
consist of C2 functions, but only C1 differentiability is required for
deriving a linear connection that defines infinitesimal differential closure.
\ If the functions are not C2, singularities can occur in second order terms,
such as curvatures (and accelerations).

\begin{center}%
%TCIMACRO{\FRAME{itbpF}{4.3223in}{2.8063in}{0in}{}{}{rmkpvlatex_fig1.wmf}%
%maintain-aspect-ratio TRUE;  display "PICT";  valid_file "F";
%width 4.3223in;  height 2.8063in;  depth 0in;  original-width 6.1358in;
%original-height 3.9686in;  cropleft "0";  croptop "1";  cropright "1";
%cropbottom "0";  filename 'rmkpvlatex_Fig1.wmf';file-properties "XNPEUR";}}}%
%BeginExpansion
{\includegraphics[
height=2.8063in,
width=4.3223in
]%
{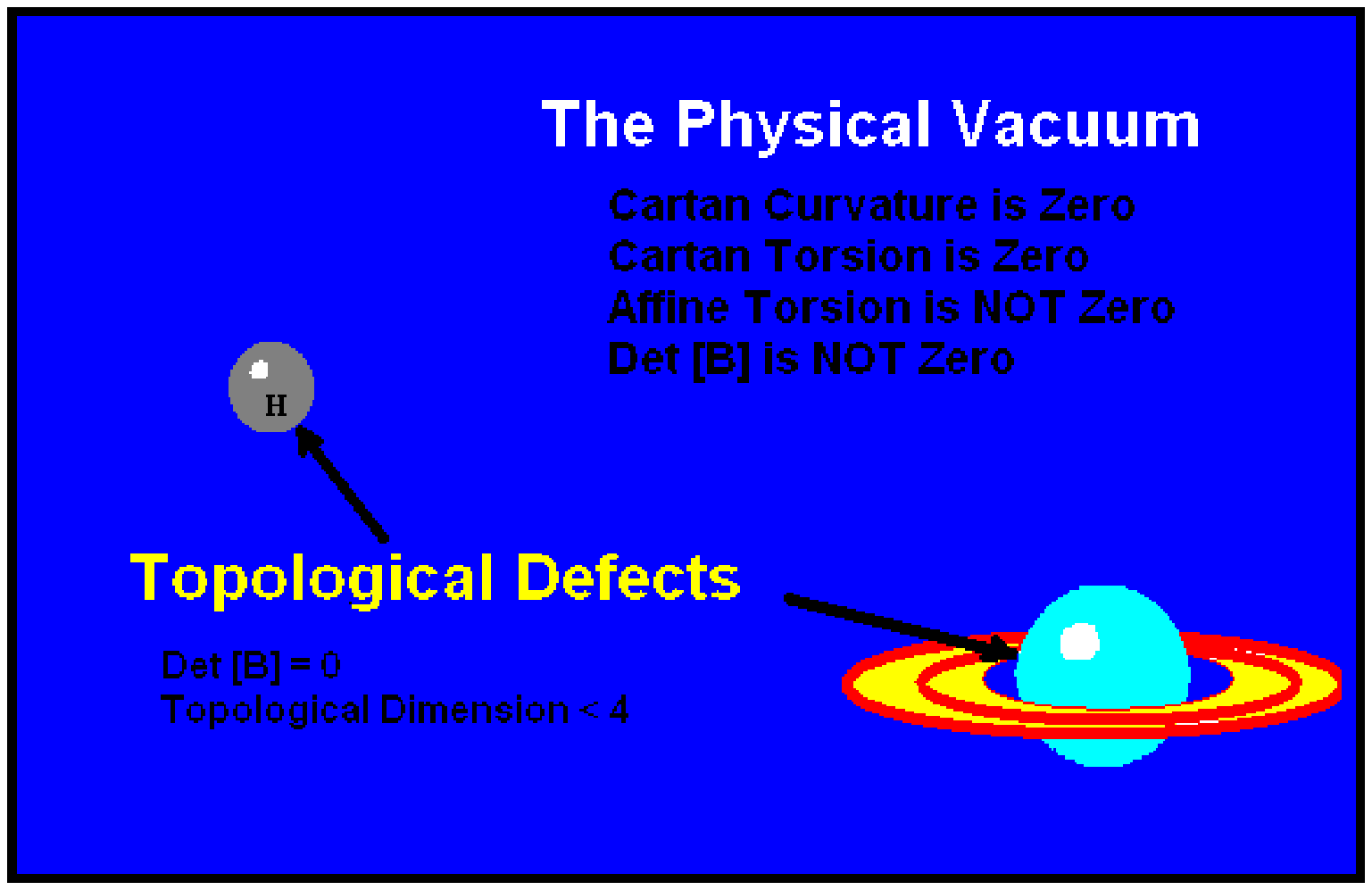}%
}%
%EndExpansion

\textbf{The 4D Cosmological Vacuum with 3D topological defects }
\end{center}

Although more complicated, the singular sets admit analysis, for example, in
terms of propagating discontinuities and topologically quantized period
integrals \cite{rmkperiods}. \ These topics will be considered in more detail
in a subsequent article.

\textbf{III}. \ It is recognized that topological coherent structures (fields,
and particles, along with fluctuations) in a "Cosmological Vacuum" can be put
into correspondence with the concepts of topological thermodynamics based on
Continuous Topological Evolution (see Vol. 1, \cite{rmklulu},
\cite{rmkcontopevol}. \ Perhaps surprising to many, topology can change
continuously in terms of processes that are not diffeomorphic. \ For example,
a blob of putty can be deformed continuously into a cylindrical rope, and then
the ends can be "pasted" together to create a non-simply connected object from
a simply connected object. \ Topological continuity requires only that the
limit points of the initial state topology be included in the closure of the
topology of the final state. \ Such continuous maps are not necessarily
invertible; it is important to remember that topology need not be conserved by
such continuous processes. \ Diffeomorphic processes require continuity of the
map and its inverse and therefor are specializations of homeomorphisms which
preserve topology. \ This observation demonstrates why tensor constraints
cannot be applied to problems of irreversible evolution and topological change
\cite{rmkretro}. \ 

In this article, the use of a Lagrange density and a variational principle to
define field equations is not of primary importance. \ The field equations are
generated with the sole constraint that they must represent thermodynamic,
continuous, topological evolution and satisfy the dynamical format of the
First Law of Thermodynamics. \ The dynamical format of the First Law of
Thermodynamics is generated by use of the Lie differential with respect to a
process direction field, $V$, acting on a system of differential forms, $\Xi$,
that encode a specific thermodynamic system. \ The method of thermodynamic
evolution can be applied to equilibrium or non-equilibrium thermodynamic
systems, to describe reversible or thermodynamically irreversible processes.

\section{Topological Structure of a Cosmological Vacuum}

\subsection{The Fundamental Postulates}

\subsubsection{Preliminaries}

Given a variety of base variables, $\{x^{a}\}$, projective geometry teaches us
that there are two kinds of maps: collineations, $\phi$, and correlations,
$\psi$. \ These concepts are readily delineated by considering the
differentials of maps that define contravariant objects and differentials of
maps that define covariant objects.%

\begin{align}
\phi &  :x^{a}\Rightarrow V^{k}(x^{a})\\
d\phi &  :\left\vert dx^{a}\right\rangle \Rightarrow\lbrack\partial
V^{k}/\partial x^{a}]\circ\left\vert dx^{a}\right\rangle =[\mathbb{B}_{a}%
^{k}]\circ\left\vert dx^{a}\right\rangle ,\\
\psi &  :x^{a}\Rightarrow A_{k}(x^{a})\\
d\psi &  :\left\vert dx^{a}\right\rangle \Rightarrow\lbrack\partial
A_{k}/\partial x^{a}]\circ\left\vert dx^{a}\right\rangle =[\mathbb{J}%
_{ka}]\circ\left\vert dx^{a}\right\rangle
\end{align}

The matrix $[\mathbb{B}_{a}^{k}]$ represents a Collineation; The matrix
$[\mathbb{J}_{ka}]$ represents a Correlation. \ In summary:%

\begin{align}
&  [\mathbb{B}_{a}^{k}]\text{ \ is a projective Collineation,}\\
&  [\mathbb{J}_{ka}]\text{ is a projective Correlation,}\\
\left[  g_{ab}\right]   &  =[\mathbb{B}]^{T}\circ\text{ }\left[  \eta\right]
\circ\lbrack\mathbb{B}\text{ \ is a symmetric Metric correlation,}\\
\left[  g_{ab}\right]   &  =\left[  g_{ba}\right]  \text{ a metrical
congruence,}\\
&  [\mathbb{J}_{ka}]\text{ is not necessarily symmetric.}%
\end{align}

In that which follows a more general idea is examined: \ the infinitesimal
collineation, $[\mathbb{B}_{a}^{k}]\circ\left\vert dx^{a}\right\rangle
=\left\vert \sigma^{k}\right\rangle $, may not be uniquely integrable. \ On a
4 dimensional variety, this means that for each 1-form $\sigma^{k}$ there are
4 possibilities:
\begin{equation}%
\begin{array}
[c]{ccc}%
\lbrack\mathbb{B}_{a}^{k}]\circ\left\vert dx^{a}\right\rangle =\left\vert
\sigma^{k}\right\rangle \ \text{but} & \text{Pfaff Sequence} & \text{PTD}\\
\sigma=d\chi\text{ \ \ \ \ \ \ \ \ \ \ \ \ \ Case 1} & d\sigma=0 & 1\\
\sigma=\phi d\chi\text{ \ \ \ \ \ \ \ \ \ \ \ Case 2} & d\sigma\neq
0,\ \sigma\symbol{94}d\sigma=0 & 2\\
\sigma=\phi d\chi+d\beta\text{ \ \ \ Case 3} & \sigma\symbol{94}d\sigma
\neq0,\ d\ \sigma\symbol{94}d\sigma=0 & 3\\
\sigma=\phi d\chi+\alpha d\beta\text{ \ Case 4} & d\ \sigma\symbol{94}%
d\sigma\neq0 & 4
\end{array}
\end{equation}
\ The Pfaff Topological Dimension (PTD) is a topological property that depends
on the functions that define the Collineation $[\mathbb{B}_{a}^{k}]$; \ The
PTD can vary from point to point in the domain $\{x^{a}\}$. \ In cases 3 and 4
the Frobenius unique integrability criteria fails, and the 1-forms are said to
be anholonomic. \ The classic examples of the failure of unique integrability
are given by porisms of envelope solutions, which are not unique, and of
characteristics, which represent discontinuities, such as an edge of
regression. \ 

\subsubsection{Postulate 1: \ The Cosmological Vacuum Field space}

The first postulate of the Cosmological vacuum, assumes the existence of a
matrix array of 0-forms (C2 functions), $\left[  \mathbb{B}\right]  =\left[
\mathbb{B}_{\operatorname{col}}^{\operatorname{row}}(x)\right]  =\left[
\mathbb{B}_{a}^{k}(x)\right]  ,$ on a 4D variety of points $\{x^{a}\}$. \ The
arbitrary matrix array of functions divides the 4D variety into two
topological regions: \ the Continuum, or field domain, where the determinants
of the matrix of functions are not zero; and the compliment of the Continuum,
or the domain of topological defects in the continuum, where the determinant
of the matrix of functions \ is zero. \ This complement of the Continuum can
be used to represent particles, condensates, wakes, solitons, and propagating
or stationary discontinuities in 4D. \ 

The Field domain is a vector space defined by the invertible matrix basis
frames, $\left[  \mathbb{B}\right]  $, \ which map a vector of exact
differentials, $\left\vert dx^{k}\right\rangle $, into a vector of exterior
differential 1-forms, $\left\vert \sigma^{k}\right\rangle $. \ The matrix,
$\left[  \mathbb{B}\right]  $, defines a Basis Frame for a vector space
constrained to those (not necessarily global) neighborhoods where the inverse
Frame, $\left[  \mathbb{B}\right]  ^{-1}$, exists:%

\begin{equation}
\text{Infinitesimal mappings: \ }\left[  \mathbb{B}\right]  \circ\left\vert
dx^{a}\right\rangle =\left\vert \sigma^{k}\right\rangle . \label{FA}%
\end{equation}
If the four 1-forms $\sigma^{k}$ are integrable, then the Basis Frame
describes "holonomic" Frames and a linear connection,
\begin{equation}
\text{Integrable mappings: \ }\left[  \mathbb{B}\right]  \circ\left\vert
X^{a}\right\rangle =\left\vert Y^{k}\right\rangle ,
\end{equation}
relating vectors of functions $\left\vert X^{a}\right\rangle $ to vectors of
functions $\left\vert Y^{k}\right\rangle $. \ If the four 1-forms $\sigma^{k}$
are NOT\ integrable, then the Basis Frame describes maps to anholonomic
1-forms "which appear naturally even in (pseudo) Riemannian geometry if
off--diagonal metrics are considered" \cite{Vacaru}.

The fundamental assumption, Eq. (\ref{FA}) is interpreted as a map of the
differential tangent vector, $\left\vert dx^{k}\right\rangle $, into
differential fibers, $\left\vert \sigma^{k}\right\rangle $, where, from
thermodynamics, the 1-forms, $\left\vert \sigma^{k}\right\rangle ,$ will have
physical dimensions of Action per unit source, $\hslash/m_{0}$. \
\begin{align}
\left[  \mathbb{B}\right]  \circ\left\vert dx^{a}\right\rangle  &  =\left\vert
\sigma^{k}\right\rangle \\
\text{Physical dimension of }\sigma^{k}  &  =(\hslash/m_{0}).
\end{align}
\ The quantity\ $(\hslash/m_{0})$\footnote{Perhaps first utilized by Onsager
in the description of superfluids} (the physical dimensions of the 1-forms
$\left\vert \sigma^{k}\right\rangle $) has the units similar to those of a
kinematic viscosity in hydrodynamics. \ Hence division of $\sigma^{k}$ by
$\hslash/m_{0}$ yields a dimensionless 1-form analogous to the classic
Reynolds number\ approach in classical hydrodynamic theory. \ The Reynolds
number idea is equivalent to the philosophy of self-similarity and homogeneity
incorporated into the Buckingham Pi theorem. \ As in the Renormalization
Group, the critical points can be determined in terms of dimensionless variables.

For an angular momentum measured in terms of characteristic mass of galaxies,
$M_{0}$, velocity, $V$, and length, $L$, the characteristic Reynolds number
can be quite large, indicating that the "fluid" is more than likely turbulent:%

\begin{equation}
\text{Rey}=VL/\mu_{B}=VL/(\hslash/M_{0}).
\end{equation}
For particles with characteristic mass, $m_{0}$, velocity $v$, and
characteristic size, $\lambda/2\pi=\hslash/m_{0}c$, the effective Reynolds
number is the ratio of velocities, $\beta=v/c<1$. \ The 1-forms, $\left\vert
\sigma^{k}\right\rangle $, can be compared directly to the electromagnetic
potentials. \ The unit "source" or "mole number" then becomes "unit charge".

\begin{claim}
The domain for which the determinant of those Basis frames that define
infinitesimal maps defines a non-global Vector space. \ It is assumed that
such regions\ will serve as the domain of the Cosmological Vacuum field. \ 
\end{claim}

The Basis Frame, $\left[  \mathbb{B}\right]  $, maps a vector array of exact
differentials into a vector array of exterior differential forms. \ Each
component of $\left\vert \sigma^{k}\right\rangle $ may or may not be exact.
\ If the components of the 1-forms are integrable in the sense of Frobenius,
$\left\vert (\sigma\symbol{94}d\sigma)^{k}\right\rangle =0$, then the
differential mappings are said to define a "trivial" bundle of "tangent
vectors". \ When the components, $\left\vert \sigma^{k}\right\rangle $, are
not uniquely integrable, the mappings are said to define a "non-trivial"
bundle of tangent vectors. \ The non-integrability, or anholonomic, feature of
the 1-forms, $\sigma^{k}$, have led to "N" (for non-linear) connections in
Finsler Spaces which can be designed to produce off-diagonal terms in metric
congruences of the Basis Frames.

The class of Basis Frames to be studied may have a determinant which is
positive definite, or which consists of either a negative domain or a positive
domain. \ In the latter case, it may be true that there exist non-unique (more
than one) solutions which can describe the "twisting of the tangent vectors".
\ It is this feature that leads to possible Chiral properties of the
Cosmological Vacuum, and enantiomorphisms of orientation.

\subsubsection{The Cosmological Vacuum Particle space}

The compliment of the Cosmological Vacuum field domain, is a singular domain,
defined as that set of points where the determinant of the Basis Frame for
infinitesimals goes to zero. \ The singular domain can have many
sub-structures. \ It is assumed that this singular domain is the realm of
topologically-coherent defect structures (such as particle condensations,
field discontinuities, open strings connecting branes, one dimensional cyclic
paths, and even Null vectors, such as Cartan Spinors). \ The concept of
topological coherence implies that these structures have recognizable
properties under deformations (where distance is not preserved), and long, if
not infinite, evolutionary lifetimes associated with solitons. \ Topological
evolution occurs as these singular objects emerge form the Cosmological Vacuum
field, or as one singular structure evolves into another. \ Thermodynamically,
irreversible processes can cause the emergence of the topological defects from
the field domain of Pfaff Topological dimension 4. \ Examples of such
continuous processes can produce the defect structures in finite time,
emulating phase changes.

\begin{claim}
\textit{The Cosmological Vacuum Particle domain is included in the compliment
of the Vector Space that defines the Cosmological Vacuum Field domain. \ It is
part of a "singular" domain where the determinant of the Basis Frame that
defines the continuum field is\ zero. \ \ \ \ }
\end{claim}

\subsubsection{Postulate 2: \ Cosmological Evolution $\approx$ Thermodynamic
Evolution}

The Cosmological Vacuum is presumed to be a thermodynamic system of exterior
differential forms, and its dynamics relative to a process, $V$, must be
described in terms of a topological realization of the First Law of
Thermodynamics. \ No other constraints of symmetry, parallel transport,
isometry, or diffeomorphic equivalences are placed on the dynamics, except to
specialize a particular problem. \ It is usual to determine dynamical
constraints by integral variational methods imposed on some Lagrange density.
\ Such methods are not assumed herein, for typically such methods do not lead
to representations of non-equilibrium systems and thermodynamically
irreversible processes.

Given any p-form, $\omega$, Cartan's magic formula expressing the exterior Lie
differential of a p-form, defines the First law as a topological statement in
terms of deRham cohomology:%

\begin{equation}
L(V)\omega=i(V)d\omega+d(i(V)\omega)=W+dU=Q.
\end{equation}
When $\omega=A$, defines 1-form of "Action per unit mole (source)", Cartan's
magic formula can be compared, explicitly, to conventional concepts associated
with (reversible or irreversible) processes, $V$, acting on (equilibrium or
non-equilibrium) thermodynamic systems (see Vol. 1 \cite{rmklulu}).%
\begin{align}
W  &  =\text{ the virtual Work 1-form,}\\
U  &  =\text{ the internal energy 0-form,}\\
Q  &  =\text{the Heat 1-form. }%
\end{align}
This method, based on Continuous Topological Evolution as generated by the Lie
differential, transcends the diffeomorphic constraints of tensor analysis.
\ For the Vector space of infinitesimal mappings, any field equations are
required to obey the rules of thermodynamic evolution, whereby the physical
system is described by a system of 1-forms, $\left\vert \sigma\right\rangle $
and the dynamics is defined with respect to a process direction field, $V$, in
in terms if the Lie differential:
\begin{equation}
L_{(V)}\left\vert \sigma\right\rangle =i(V)d\left\vert \sigma\right\rangle
+d(i(V)\left\vert \sigma\right\rangle )=W+dU=Q.
\end{equation}

\subsection{Beyond Diffeomorphic Equivalence}

\subsubsection{Matrix Methods}

The idea is to use matrix representations of differential forms and the
exterior matrix product to replicate and expand the ideas generated by the
ubiquitous Tensor analysis. \ Diffeomorphic equivalence - used to define
tensors - eliminates the possibility of topological change. \ Matrix methods,
with elements composed of exterior differential forms, do not impose the
constraint of diffeomorphic equivalence. \ 

The fundamental assumption is that the Basis Frames, which lead to the Cartan
connection, are representations of infinitesimal mappings. \ 

The formalism includes matrices of 0-forms (to defined Basis Frames and vector
spaces), matrices of 1-forms (used to define a Connection) matrices of 2-forms
(used in the definition of Curvature) as well as matrices of 3-forms (used to
define Bianchi identities). \ 

\begin{enumerate}
\item Matrices of 0-forms: \ Basis Frames $[\mathbb{B}]$ and metric $[g]$

\item Matrices of 1-forms: \ Connections, $[\mathbb{C]}_{\text{Cartan based on
[B]}}$, $[\Gamma]_{\text{Christoffel on [g]}}\,$, $\ [\mathbb{T}%
]_{\text{Residue}}=[\mathbb{C}]-[\Gamma].$

\item Matrices of 2-forms: \ Curvatures, $[\Theta]_{\text{based on [C]}},$
$[\Phi\mathbb{]}_{\text{based on [}\Gamma]}$, $[\Sigma\mathbb{]}_{\text{based
on [}\mathbb{T}\text{]}}.$
\end{enumerate}

\ In addition, the formalism utilizes a number of vector arrays of p-forms
generated from the fundamental assumption, Eq.(\ref{FA}) :

\begin{enumerate}
\item Vectors of pair 1-forms are used to define tangent spaces and Action per
unit source,
\begin{equation}
\left\vert \sigma^{k}\right\rangle \approx e/m_{0}\left\vert A^{k}%
\right\rangle ;
\end{equation}
$\ $the potentials.

\item Vectors of exact pair 2-forms are used to define Field Intensity
2-forms,
\begin{equation}
F^{k}=\left\vert d\sigma^{k}\right\rangle \approx e/m_{0}\left\vert
dA^{k}\right\rangle .
\end{equation}

\item Vectors of pair 3-forms are used to define the Topological Torsion
generated by
\begin{equation}
\left\vert \sigma\symbol{94}d\sigma\right\rangle \approx(e/m)^{2}\left\vert
A\symbol{94}dA\right\rangle =(e/m)^{2}\left\vert A\symbol{94}F\right\rangle
=i(\mathbf{T}_{4})\Omega_{4}.
\end{equation}

\item Vectors of pair 4-forms are used to define Topological Parity and bulk
dissipation,
\begin{equation}
\left\vert d\sigma\symbol{94}d\sigma\right\rangle \approx(e/m)^{2}\left\vert
dA\symbol{94}dA\right\rangle =(e/m)^{2}\left\vert F\symbol{94}F\right\rangle
=-2(\mathbf{E\circ B})\Omega_{4}=PoincareII.
\end{equation}

\item Vectors of impair 2-form densities are used to define the Affine Torsion
fields, $[\mathbb{C}]\symbol{94}\left\vert dx^{n}\right\rangle $, of
hydrodynamics, as $\left\vert G^{k}\right\rangle $. \ 

\item Vectors of impair 3-form densities are used to define the closed 3-form
currents of electromagnetism, $\left\vert J^{k}\right\rangle =\left\vert
dG^{k}\right\rangle $

\item Vectors of impair 3-form pseudoscalar densities are used to define
Topological Spin in electromagnetism, \ $\left\vert A\symbol{94}G\right\rangle
.$

\item Vectors of impair 4-form densities are used to define the Poincare
Lagrange pseudoscalar density of electromagnetism,
\begin{equation}
PoincareI=d(A\symbol{94}G)=(F\symbol{94}G-A\symbol{94}J).
\end{equation}

\item Vectors of pair 2-form densities are used to define the Affine Torsion
fields, $[\mathbb{C}]\symbol{94}\left\vert dx^{n}\right\rangle $, of
hydrodynamics, as $\left\vert \mathfrak{T}^{k}\right\rangle $. \ 

\item Vectors of pair 3-form densities are used to define the Torsion Spin
3-forms, \ $\left\vert \sigma\symbol{94}\mathfrak{T}\right\rangle .$ \ 
\end{enumerate}

Each of the formulas are vector indexed k to represent the elements of the
Vectors so described by the assumptions of the Cosmological Vacuum. \ 

Using tensor methods (that impose a Connection arbitrarily) it is easy to
overlook the fact that for every Basis Frame, $\left[  \mathbb{B}\right]  $,
of an infinitesimal vector space, there exists both a Right Cartan Connection,
$[\mathbb{C}]$, and a Left Cartan Connection, $[\Delta]$. $\ $\ Each
Connection is a matrix of 1-forms. \ %

\begin{align}
d[\mathbb{B}]  &  =[\mathbb{B}][\mathbb{C}]=[\Delta][\mathbb{B}],\\
\lbrack\mathbb{C}]  &  =[\mathbb{B}]^{-1}[\Delta][\mathbb{B}].
\end{align}
The two connections are related by a similarity transformation, which
preserves their symmetric properties, but does not exhibit the antisymmetry
and chiral properties associated with torsion. \ The matrix methods, utilized
below, correct these deficiencies\textit{, and will demonstrate Chirality
effects associated with both matrices of 3-forms and with matrices of
1-forms.}

\subsection{Constructive Results for any Basis Frame}

By applying algebraic and exterior differential processes. developed by E.
Cartan, to each element, $\left[  \mathbb{B}\right]  $, of given equivalence
class of Basis Frames of C2 functions, the following concepts are
\textbf{derived, not postulated}. \ It is possible to encode these results in
terms of existence theorems, but they are best demonstrated by constructive
proofs. \ Starting from the fundamental assumption, eq. (\ref{FA} ), the
following constructions are possible:

A "flat" Right Cartan Connection as a matrix of 1-forms, $[\mathbb{C]}$, can
be derived, leading to structural equations of curvature 2-forms and Cartan
Torsion 2-forms, both of which are zero relative to the Cartan Connection.
\ However, the Cartan Connection can support the concept of non-zero "Affine
Torsion". \ The Right Cartan Connection, $[\mathbb{C}]$, is algebraically
compatible with the Basis Frame, and the Vector space that it defines. \ From
the identity $\left[  \mathbb{B}\right]  \circ\left[  \mathbb{B}\right]
^{-1}=\left[  \mathbb{I}\right]  ,$ use exterior differentiation to
\textit{derive} the (right) Cartan Connection $\left[  \mathbb{C}\right]  $ as
a matrix of 1-forms: \ \
\begin{align}
\text{Right Cartan Connection }\text{: }  &  \left[  \mathbb{C}\right]  \text{
}\nonumber\\
d\left[  \mathbb{B}\right]  \circ\left[  \mathbb{B}\right]  ^{-1}+\left[
\mathbb{B}\right]  \circ d\left[  \mathbb{B}\right]  ^{-1}  &  =d\left[
\mathbb{I}\right]  =0\\
\text{ \ hence }d\left[  \mathbb{B}\right]   &  =\left[  \mathbb{B}\right]
\circ\left[  \mathbb{C}\right]  ,\\
\text{\ where\ }\left[  \mathbb{C}\right]   &  =-d\left[  \mathbb{B}\right]
^{-1}\circ\left[  \mathbb{B}\right] \\
&  =+\left[  \mathbb{B}\right]  ^{-1}\circ d\left[  \mathbb{B}\right] \\
&  =[C_{a\ m}^{b}(y)dy^{m}]=[C_{\text{column }m}^{\text{row}}(y)dy^{m}].
\end{align}
The Connection leads to the idea of differential closure, in the sense that
the differential of any column vector of the Basis Frame is (at a point)\ a
linear combination of the column vectors that make up the Basis Frame.

It is also possible to construct a Left Cartan Connection matrix of 2-forms,
$[\Delta]$, relative to the Frame Matrix, $[\mathbb{B}]$\ such that:%
\begin{align}
\text{Left}  &  :\text{ Cartan Connection }\left[  \Delta\right]
\text{\ \ }\\
d\left[  \mathbb{B}\right]   &  =\left[  \Delta\right]  \circ\left[
\mathbb{B}\right]  ,\\
\left[  \Delta\right]   &  =-\left[  \mathbb{B}\right]  \circ d\left[
\mathbb{B}\right]  ^{-1}\\
&  =+d\left[  \mathbb{B}\right]  \circ\left[  \mathbb{B}\right]
^{-1}=-[\mathbb{C}].
\end{align}
The coefficients that make up the matrix of 1-forms, $\left[  \mathbb{\Delta
}\right]  ,$ can be associated with what have been called the Weitzenboch
Connection coefficients..

The Right and Left Cartan connections are not (usually) identical. \ They are
equivalent in terms of the similarity transformation: \
\begin{equation}
\left[  \mathbb{C}\right]  =\left[  \mathbb{B}\right]  ^{-1}\circ\left[
\Delta\right]  \circ\left[  \mathbb{B}\right]  .
\end{equation}
The left Cartan Connection, in general, is \textit{not} the same as the
transpose of the right Cartan Connection. \ 

Also note that inverse matrix also enjoys differential closure properties.
\begin{equation}
d\left[  \mathbb{B}\right]  ^{-1}=\left[  \mathbb{B}\right]  ^{-1}\circ\left[
-\Delta\right]  =\left[  -\mathbb{C}\right]  \circ\left[  \mathbb{B}\right]
^{-1}.
\end{equation}

\subsubsection{Orthonormal Frames - a special case}

If the Basis Frame is orthogonal such that the transpose $\left[
\mathbb{B}\right]  ^{T}$ is equal to the inverse $\left[  \mathbb{B}\right]
^{-1}$, then the Cartan matrix of Connection 1-forms is anti-symmetric.%

\begin{align}
\text{Orthonormal Tetrads}  &  \text{: }\left[  \mathbb{B}\right]
^{T}=\left[  \mathbb{B}\right]  ^{-1},\text{ det}[\mathbb{B}]=1\\
\lbrack\mathbb{C}]  &  =-[\mathbb{C}]^{T}\text{.}%
\end{align}
The determinant of the orthogonal group can be $\pm1$, but the orthonormal
group implies that the determinant is positive definite. \ At this level,
there is no need to presume that the Basis Frame is an element of the
orthonormal group, but this choice is often made as a particular "gauge"
constraint in metrical theories. \ The idea of gauge can be associated with
the constraint that the Basis Frame $[\mathbb{B}]$ be an element of a
particular subgroup of the general linear group of transformations.

It is somewhat surprising to me, but quite often it is claimed that this
anti-symmetric Connection is a "Spin connection". \ A different point of view
is taken is this article. \ 

\subsubsection{Symmetric Congruences - the metric}

A symmetric quadratic congruence of functions, relative to a signature matrix,
$[\eta]$, can be deduced by algebraic methods from Basis Frame. \ This
symmetric congruence can play the role of a metric of 0-forms, $\left[
g\right]  $, compatible with the Basis Frame:\
\begin{equation}
\left[  g\right]  =[\mathbb{B}]^{T}\circ\lbrack\eta]\circ\left[
\mathbb{B}\right]  .
\end{equation}
The quadratic form, $\left[  g\right]  $, can be used to generate a
Christoffel Connection, $\left[  \Gamma(g)\right]  $, not equal to the Cartan
Connection. \ The Christoffel computation is metric dependent and depends upon
the partial derivatives of the symmetric congruence. \ The Christoffel
Connection is not (necessarily)\ flat, and can produce Cartan curvature
equations of structure which are not zero. \ The Christoffel Connection will
not produce Affine Torsion. \ For a given Basis Frame, $\left[  \mathbb{B}%
\right]  $, the Cartan Connection, $\left[  \mathbb{C}\right]  $ can be
decomposed into the sum of the Christoffel Connection, $\left[  \Gamma
(g)\right]  ,$ and a residue $[\mathbb{T}]$.
\begin{equation}
\text{Connection Decomposition formula }\left[  \mathbb{C}\right]  =\left[
\Gamma(g)\right]  +[\mathbb{T}]
\end{equation}
If the metric is diagonal, then it is possible to construct a special
compatible Basis Frame which will generate the diagonal metric as symmetric
congruence (see example 1, below).

\subsubsection{Matrices of Curvature 2-forms}

It is common to define the Cartan matrix of curvature 2-forms, $\left[
\mathbf{\Phi}\right]  $, for any arbitrary\footnote{The arbitrary connection
$[\Gamma]$ is not necessarily a Cartan Connection, $[\mathbb{C}]$, defined by
a Basis Frame $[\mathbb{B}]$.} Connection $[\Gamma]$, in terms of the formula:%
\begin{equation}
\text{ Matrix of Curvature 2-forms \ }\left[  \mathbf{\Phi}\right]
=[\Gamma]\symbol{94}[\Gamma]+d[\Gamma].
\end{equation}
This equation is said to define Cartan's second equation of structure. \ Note
that exterior differentiation of the Cartan structure matrix of curvature
2-forms is equivalent to a Bianchi identity:%
\begin{align}
\left[  d\mathbf{\Phi}\right]  +\left[  d\Gamma\right]  \symbol{94}\left[
\Gamma\right]  -\left[  \Gamma\right]  \symbol{94}\left[  d\Gamma\right]   &
=\\
\left[  d\mathbf{\Phi}\right]  +\left[  \mathbf{\Phi}\right]  \symbol{94}%
\left[  \Gamma\right]  -\left[  \Gamma\right]  \symbol{94}\left[
\mathbf{\Phi}\right]   &  \Rightarrow0.
\end{align}
This concept of a Bianchi identity is valid for all forms of the Cartan
structure equations. \ The Bianchi statements are essentially definitions of
cohomology, in that the difference between two non-exact p-forms is equal to a
perfect differential (an exterior differential system). \ In this case the
Bianchi identity describes the cohomology established by two matrices of
3-forms, $\left[  J2\right]  -\left[  J1\right]  $.%
\begin{align}
\left[  d\mathbf{\Phi}\right]   &  =\left[  J2\right]  -\left[  J1\right]  ,\\
\text{where }\left[  J1\right]   &  =\left[  d\Gamma\right]  \symbol{94}%
\left[  \Gamma\right] \\
\text{and }\left[  J2\right]   &  =\left[  \Gamma\right]  \symbol{94}\left[
d\Gamma\right]  .
\end{align}
These formulas can be interpreted in terms of Chiral properties of the
Structure when the closed integrals of each 3-form is not zero, but the closed
integrals have equal and opposite signs. \ Note that is possible that each
3-form is not exact, but their difference is always exact. \ Such is the stuff
of deRham cohomology theory.

In certain cases the two Chiral species are equivalent, for which the exterior
differential of the Curvature vanishes. \ In other cases, the two Chiral
species are NOT\ equal, and the exterior differential of the Curvature is not
zero. \ It is remarkable that in all cases the exterior differential of each
term, though possibly different, cancel each other. \ These 3-forms represent
Vectors of 3-forms, which may or may not have Zero divergence, but in all
non-zero cases the divergence of $[RH_{3}]$ is equal and opposite in sign to
the other, $[LH_{3}]$.

\subsection{Constructions for Basis Frames of infinitesimals}

\subsubsection{Infinitesimal Neighborhoods, Affine Torsion}

Now (IMO) it is most remarkable that the exterior differential of Eq.
(\ref{FA}) leads to the equations,%

\begin{align}
\lbrack\mathbb{B}_{m}^{k}]\circ\lbrack\mathbb{C}]\symbol{94}\left\vert
dx^{m}\right\rangle  &  =\left\vert F^{k}\right\rangle ,\\
d[\mathbb{B}_{m}^{k}]  &  =[\mathbb{B}_{m}^{k}]\circ\lbrack\mathbb{C}],
\end{align}
where $[\mathbb{C}]$ is the right Cartan matrix of connection 1-forms,
relative to the matrix of Basis Functions. \ The fascinating result is that
the formula,%

\begin{equation}
\lbrack\mathbb{C}]\symbol{94}\left\vert dx^{m}\right\rangle =\left\vert
\text{Affine Torsion 2-forms}\right\rangle ,
\end{equation}
precisely defines the classic vector of Affine Torsion 2-forms!!! \ The Cartan
Curvature 2-forms, $\Theta=\{d[\mathbb{C}]+[\mathbb{C}]\symbol{94}%
[\mathbb{C}]\}$ are zero, so the space is "flat", but Affine Torsion persists. \ 

\subsubsection{Pair differential 2-form densities (mass)}

Suppose that the determinant of Basis Frame functions is positive definite.
\ Then concepts of orientation are not important and there exists a "center of
symmetry". \ In such cases, the vector of Affine Torsion 2-forms consists of
"pair" exterior differential form densities with integrations that do not
depend upon orientation. \ Such cases will utilize the notation,%
\begin{equation}
\text{Affine Torsion (pair) \ }[\mathbb{C}]\symbol{94}\left\vert
dx^{m}\right\rangle =\left\vert \mathfrak{T}^{m}\right\rangle .
\end{equation}
The period integrals of such forms are always positive, emulating the physical
properties of mass.%

\begin{equation}
M=\int\int_{2dcycle}\mathfrak{T}>0
\end{equation}

\subsubsection{Impair differential 2-form densities (charge)}

If the determinant of the Basis Frame is negative, then the vector of Affine
Torsion 2-forms will be given the notation:%
\begin{equation}
\text{ Affine Torsion (impair) }\left[  \mathbb{C}\right]  \symbol{94}%
\left\vert dx^{m}\right\rangle =\left\vert G^{m}\right\rangle .
\end{equation}
The 2-forms $\left\vert G^{m}\right\rangle $ are sensitive to the orientation
and are impair differential form densities. \ It is this sensitivity to
orientation of the volume element that permits the period integrals of
$\left\vert G^{m}\right\rangle $ to have both positive and negative values. \ %

\begin{equation}
Q=\int\int_{2dcycle}G\gtrless0.
\end{equation}

The two equations relating Affine Torsion 2-forms to field intensities
$\left\vert F^{k}\right\rangle $, appear to be the equivalent of a
constitutive equation mapping, mapping the Affine Torsion 2-forms $\left[
\mathbb{C}\right]  \symbol{94}\left\vert dx^{m}\right\rangle $ into field
intensities $\left\vert F^{k}\right\rangle .$%
\begin{align}
\lbrack\mathbb{B}_{m}^{k}]\circ\left\vert \mathfrak{T}^{m}\right\rangle  &
=\left\vert F^{k}\right\rangle ,\\
\left\vert \mathfrak{T}^{m}\right\rangle  &  =\text{pair density 2-forms,
or}\\
\lbrack\mathbb{B}_{m}^{k}]\circ\left\vert G^{m}\right\rangle  &  =\left\vert
F^{k}\right\rangle ,\\
\left\vert G^{m}\right\rangle  &  =\text{impair density 2-forms,}%
\end{align}

\ In the first case, the unit source will only have positive values (of mass),
and in the second case the unit source can have both positive and negative
values (of charge). \ 

\begin{conjecture}
I contend if det($[\mathbb{B}_{m}^{k}])>0$ then $\left[  \mathbb{C}\right]
\symbol{94}\left\vert dx^{m}\right\rangle \Rightarrow$ $\left\vert
\mathfrak{T}^{m}\right\rangle $ is pair and has periods that represent mass.
\ If det($[\mathbb{B}_{m}^{k}])<0$ then $\left[  \mathbb{C}\right]
\symbol{94}\left\vert dx^{m}\right\rangle \Rightarrow$ $\left\vert
G^{m}\right\rangle $ is impair and has periods that represent charge. \ 
\end{conjecture}

\subsubsection{Curvature 2-forms}

Note, however, that a second exterior differentiation of the infinitesimal
constraint equation, eq. (\ref{FA}), yields:
\begin{align}
d\{\left[  \mathbb{B}\right]  \circ\left[  \mathbb{C}\right]  \symbol{94}%
\left\vert dx^{k}\right\rangle \}  &  =\left[  \mathbb{B}\right]
\circ\{\left[  \mathbb{C}\right]  \symbol{94}\left[  \mathbb{C}\right]
+d[\mathbb{C}]\}\symbol{94}\left\vert dx^{k}\right\rangle \\
&  =\left[  \mathbb{B}\right]  \circ\left[  \Theta\right]  \symbol{94}%
\left\vert dx^{k}\right\rangle =\left\vert dd\sigma^{k}\right\rangle \approx
e/m_{0}\ \left\vert ddA^{k}\right\rangle =0.
\end{align}
Hence the matrix of Curvature 2-forms for a Right Cartan Connection of
infinitesimals is zero. \ The Cartan equations of structure, based on the
Right Cartan Connection, $[\mathbb{C}]$, indicate that the Basis Frame of
infinitesimals defines a vector space of zero curvature, but non-zero Affine
torsion. \ This vector space is defined as the vector space of the fields of
the Cosmological Vacuum.

On the other hand, the Christoffel connection, $\left[  \Gamma(g)\right]  $,
based on a symmetric congruence always produces a symmetric connection, for
which the structural equations permit non-zero curvature, but zero Affine
torsion. \ 

If the vector space defined by the infinitesimal mappings, $\left[
\mathbb{B}\right]  \circ\left\vert dx^{k}\right\rangle =$ $\left\vert
\sigma^{m}\right\rangle $ is constrained by a metric field, $[g]$, of
congruent mappings, then the Christoffel Connection, $\left[  \Gamma
(g)\right]  $, can be generated from the metric coefficients. \ \ The Cartan
Connection relative to a vector space of infinitesimals can be decomposed into
two parts,%

\begin{equation}
\lbrack\mathbb{C]}=[\Gamma]+[\mathbb{T}].
\end{equation}
This decomposition will permit the study of how mass and gravity (due to
$[\Gamma]$) can be influenced by Affine Torsion (generated by $[\mathbb{T}]$).

\subsubsection{Vectors of 2-forms and 3-forms}

\paragraph{Intensities}

In both the hydrodynamic case and the electrodynamic case, the system of
intensity 2-forms, $\left\vert dA^{b}\right\rangle =m_{0}/e\ \left\vert
d\sigma^{k}\right\rangle ,$ are formally equivalent to the Maxwell-Faraday
equations of field intensities:%
\begin{align}
\left\vert dA^{b}\right\rangle  &  =\left\vert F^{b}\right\rangle \ \text{a
vector of 2-forms,}\\
\left\vert dF^{b}\right\rangle  &  \Rightarrow0,\ \ \text{ a vector of
3-forms,}\\
\text{or }  &  \text{: the Maxwell Faraday PDE,s.}%
\end{align}
Each of the exact 2-forms $\left\vert F^{b}\right\rangle $ in 4D will generate
a Symplectic manifold, if\ they are of maximal rank, $F\symbol{94}F\neq0.$
\ The notations and equations are identical to within a constant factor. \ It
is the 1-forms and the intensities that lead to the different topological
structures, and so the EM notation will be used for both the electromagnetic
and the hydrodynamic problems.

\paragraph{Currents}

In the electromagnetic case, the exterior differential of the impair 2-form,
$G$, leads to the concept of a conserved charge-current density. \ The
resulting equations have the format of the Maxwell-Ampere PDE,s. \ %

\begin{align}
\left\vert J_{em}^{k}\right\rangle  &  =d\left\vert G^{k}\right\rangle ,\\
d\left\vert J_{em}^{k}\right\rangle  &  =0.
\end{align}

In hydrodynamic theory it is most remarkable that the exterior differential of
the vector density of Affine Torsion 2-forms, $\left\vert \mathfrak{T}%
\right\rangle $, leads to a vector of closed "fluid mass"-current 3-forms:%

\begin{align}
\left\vert J_{fluid}^{k}\right\rangle  &  =d\left\vert \mathfrak{T}%
\right\rangle =[d\mathbb{C}]\symbol{94}\left\vert dx\right\rangle
=-[\mathbb{C}]\symbol{94}[\mathbb{C}]\symbol{94}\left\vert dx\right\rangle ,\\
d\left\vert J_{fluid}^{k}\right\rangle  &  =0.
\end{align}
This result is similar, but not identical to to the "charge"-current 3-form
density of electromagnetism.

The mass-current is pair density 3-form, but the charge-current density 3-form
is impair. \ In electromagnetic theory, the excitation 2-forms, \ $\left\vert
G^{b}\right\rangle $, are not exact, and the equations that lead to the closed
3-form densities are defined as the Maxwell Ampere equations. \ In domains
where $d\left\vert J_{em}^{k}\right\rangle =0$, closed integrals of
$\left\vert G^{b}\right\rangle $ can be of opposite sign. \ For the closed
"mass currents", $d\left\vert J_{fluid}^{k}\right\rangle =0$, the 2-forms of
Affine Torsion, $\left\vert \mathfrak{T}\right\rangle $, need not be exact,
but closed integrals in domains where the mass-current vanishes, $\left\vert
J_{fluid}^{k}\right\rangle =0$, are now of the same sign: \ mass is positive
definite, charge is not.\ 

The 4-component structure can lead to serious algebraic difficulties, best
overcome with a symbolic math processor, such as Maple. \ In the second part
of this article, a series of Maple programs are presented offering the details
of many examples. \ If all but one of the four 1-form components of
$\left\vert A^{k}\right\rangle $ are closed, then the formalism encodes the
topological theory of electromagnetism. \ Several examples based on Particle
Affine, Wave-Affine, and the Hopf map are presented in Part II all have this
simplistic property, and deserve closed study. \ Moreover, if the totality of
the four 1-form components of $\left\vert \sigma^{k}\right\rangle $ are not
closed, the same starting point encodes the fields that are utilized by Yang
Mills theory. \ Each of these specializations is a topological refinement.

\paragraph{Spin 3-forms}

The next thing to investigate is the concept of Spin 3-forms. Construct the
two spin 3-form densities, and the 3-form of Topological Torsion,%

\begin{align}
\text{Pair Spin 3-form densities \ }\left\vert A^{m}\symbol{94}\mathfrak{T}%
^{m}\right\rangle  &  =\left\vert A^{m}\symbol{94}[\mathbb{B}_{m}^{k}%
]^{-1}F^{k}\right\rangle ,\\
\text{Impair Spin 3-form densities \ }\left\vert A^{m}\symbol{94}%
G^{m}\right\rangle  &  =\left\vert A^{m}\symbol{94}[\mathbb{B}_{m}^{k}%
]^{-1}F^{k}\right\rangle \\
\text{Topological Torsion 3-form \ }i(\mathbf{T}_{4})\Omega_{n}  &
=\left\vert A^{m}\symbol{94}F^{m}\right\rangle
\end{align}

\paragraph{Topological Spin and the First Poincare 4-form}

\ In electromagnetic format, it is possible to construct a vector of impair
density 3-forms, $\left\vert S^{m}\right\rangle $. \ Each component, $S^{m}$,
will define a value for the Topological Spin generated by
each\footnote{Realize that each 1-form, $\sigma^{m}$, can generate a distinct
topological structure. \ Hence there are 4 possibly differnent topologies
imposed upon 4D Space-Time simultaneously.} of the 1-forms, $A^{m}$. \ The
4-divergence of each component yields the first Poincare function of (twice)
the Field Lagrangian density minus the Field Interaction energy
\cite{rmkintrinsic}.
\begin{align}
\text{Topological Spin 3-forms }  &  \text{=}\text{ }\left\vert S^{m}%
\right\rangle =\ \left\vert A^{m}\symbol{94}G^{m}\right\rangle _{em},\text{
(no sum)}\\
\text{Physical dimensions of }S^{m}  &  =(\hslash)\ \ (\text{angular
momentum)}\\
\text{ Poincare I 4-form \ }\left\vert dS^{m}\right\rangle  &  =\{\left\vert
F^{m}\symbol{94}G^{m}\right\rangle -\left\vert A^{m}\symbol{94}dG^{m}%
\right\rangle \}=\left\vert \mathcal{L}^{m}\Omega\right\rangle ,\\
\text{The 4-form coefficient of }dS^{m}  &  =\text{Lagrange Action,
}\mathcal{L}^{m},\\
\text{In EM format }\mathcal{L}^{m}  &  =\{(\mathbf{B\circ H-D\circ E)-(A\circ
J-}\rho\phi)\}^{m}%
\end{align}
\qquad\qquad

For the hydrodynamic case, it is possible to construct a vector of pair
density 3-forms, $\left\vert \mathcal{S}^{m}\right\rangle $. \ In fluid
format, each component, $\mathcal{S}^{m}$, will define a value for the
Topological Spin generated by each\footnote{Realize that each 1-form,
$\sigma^{m}$, can generate a distinct topological structure. \ Hence there are
4 possibly differnent topologies imposed upon 4D Space-Time simultaneously.}
of the 1-forms, $\sigma^{m}$. \ The 4-divergence of each component yields the
first Poincare function of (twice) the Field Lagrangian density minus the
Field Interaction energy \cite{rmkintrinsic}.
\begin{align}
\text{Topological Spin 3-forms }  &  \text{=}\text{ }\left\vert \mathcal{S}%
^{m}\right\rangle =\left\vert \sigma^{m}\symbol{94}\mathfrak{T}^{m}%
\right\rangle _{fluid}\ \ \ \text{(no sum)}\\
\text{Physical dimensions of }\mathcal{S}^{m}  &  =(\hslash)\ \ (\text{angular
momentum)}\\
\text{ Poincare I 4-form \ }\left\vert d\mathcal{S}^{m}\right\rangle  &
=\{\left\vert d\sigma^{m}\symbol{94}\mathfrak{T}^{m}\right\rangle -\left\vert
\sigma^{m}\symbol{94}d\mathfrak{T}^{m}\right\rangle \}=\left\vert
\mathcal{L}^{m}\Omega\right\rangle ,\\
\text{The 4-form coefficient of }d\mathcal{S}^{m}  &  =\text{Lagrange Action,
}\mathcal{L}^{m}%
\end{align}
Note that the existence of Topological Spin depends upon a non-zero value of
the vector of Affine Torsion 2-forms. \ In this sense, Affine Torsion is a
necessary, but not a sufficient condition for the existence of Spin. \ When
the Poincare density, $\mathcal{L}$, vanishes, then the Spin 3-form is closed.
\ The 3-dimensional integrals of the closed Spin 3-forms lead to deRham period
integrals and topological Quantization of the Spin. \ \ 

\paragraph{Topological Torsion and the Second Poincare 4-form}

It is possible to construct the vector of 3-forms that represent Topological
Torsion, $\left\vert H^{m}\right\rangle $, and its divergence, defined as
Topological Parity, $\left\vert K^{m}\right\rangle $. \ $\left\vert
K^{m}\right\rangle $ is a vector array of 4-forms, with coefficients that
define the second Poincare function. \ The physical dimensions of $K$ are
proportional to the square of the kinematic viscosity defined by
\begin{align}
\text{Topological Torsion 3-forms }  &  \text{=}\text{ }\left\vert
H^{m}\right\rangle =\left\vert \sigma^{m}\symbol{94}d\sigma^{m}\right\rangle
\approx(e/m_{0})^{2}\ \left\vert A^{m}\symbol{94}F^{m}\right\rangle ,\text{
}\\
\left\vert K^{m}\right\rangle  &  =d\left\vert H^{m}\right\rangle =\left\vert
d\sigma^{m}\symbol{94}d\sigma^{m}\right\rangle \approx(e/m_{0})^{2}%
\ \left\vert F^{m}\symbol{94}F^{m}\right\rangle ,\\
\text{The Physical dimensions of each }K^{m}  &  =(\hslash/m_{0})^{2}%
=(\hslash)(\hslash/m_{0}^{2})=(\hslash)(\hslash/e_{0}^{2})(e/m_{0})^{2}%
\end{align}
Note that the physical dimensions of each Topological Torsion 3-form have the
physical dimensions of a kinematic viscosity squared:%
\begin{equation}
(\hslash/m_{0})^{2}=\mu_{B}^{2}.
\end{equation}
This kinematic velocity is not the same as a "shear" viscosity, but is more
sensibly described as a "bulk viscosity" \cite{Eckart}. \ Further note the
natural inclusion of the Hall impedance, $(\hslash/e_{0}^{2})$, and its
relationship to the "bulk viscosity", $(\hslash/m_{0})^{2}.$
\begin{align}
\text{ Poincare II 4-form }\left\vert K^{m}\right\rangle  &  =(e/m_{0}%
)^{2}\ \left\vert F^{m}\symbol{94}F^{m}\ \right\rangle \ ,\\
\text{In EM format, for each m, }K^{m}  &  =(e/m_{0})^{2}\{-2(\mathbf{B\circ
E)}^{m}\}\Omega
\end{align}

\paragraph{Chiral 1-forms}

Perhaps the most intriguing thing about Einstein's formula\footnote{The
Einstein formula uses the contracted Ricci tensor and the Ricci scalar to
construct
\par
$G_{m}^{c}=g^{cb}R_{bma}^{a}-\delta_{m}^{c}R/2$} (which is used to define the
concept of gravity in terms of the deformation of space-time) is that the
Einstein tensor \textsf{G}$_{m}^{c}$ is deduced from contractions of the
Riemann curvature tensor that produce the "Ricci tensor". \ \ The formulation
of the Einstein tensor, \textsf{G}$_{m}^{c}$, is said to have "Zero
divergence". \ The Riemann curvature tensor depends only upon the connection,
but the formulation of the Einstein tensor also implicates the metric
structure (to raise and lower the indices), as well. \ Why Einstein chose the
particular tensor combination to define gravity is not obvious, for there are
many structures that have a zero exterior derivative.

As demonstrated below, it is possible to generate two matrices of 1-forms,
with a sum equal to zero, but with a difference which is not zero, but is
always exact. \ Some of the properties of the Einstein tensor can be
replicated by a vector of 1-forms that has zero exterior derivative, and is
constructed directly from the Connection without recourse to the metric. \ The
formulas demonstrate by construction the concept of a vector of 1-forms, that
always has a ZERO\ exterior differential, and yet can be composed of
components, that may exhibit Chiral properties.

By exterior differentiation of the fundamental assumption, eq (\ref{FA}),
\begin{align}
d\{\left[  \mathbb{B}\right]  \circ\left\vert dx^{k}\right\rangle \}  &
=d\left[  \mathbb{B}\right]  \circ\left\vert dx^{k}\right\rangle =\left\vert
d\sigma^{k}\right\rangle ,\\
\{\left[  \mathbb{B}\right]  \circ\mathbb{C}\symbol{94}\left\vert
dx^{k}\right\rangle \}  &  =\left[  \mathbb{B}\right]  \circ\left\vert
\mathfrak{T}^{k}\right\rangle =\left\vert d\sigma^{k}\right\rangle
\end{align}
Recall that the vector of 2-forms, $\left\vert \mathfrak{T}^{k}\right\rangle
=\left[  \mathbb{C}\right]  \symbol{94}\left\vert dx^{k}\right\rangle $,
defines the vector of Affine Torsion 2-forms, $\left\vert \mathfrak{T}%
^{k}\right\rangle $, relative to the Right Cartan connection, $\left[
\mathbb{C}\right]  $.

It is to be noted that the inverse Basis Frame, $\left[  \mathbb{B}\right]
^{-1}$, maps differential forms into exact differentials,%

\begin{equation}
\left\vert dx^{k}\right\rangle =\left[  \mathbb{B}\right]  ^{-1}%
\circ\left\vert \sigma^{k}\right\rangle .
\end{equation}
Exterior differentiation leads to the expression,
\begin{equation}
\left[  -\Delta\right]  \circ\left\vert \sigma^{k}\right\rangle =[-\Delta
]\circ\left[  \mathbb{B}\right]  \circ\left\vert dx^{k}\right\rangle
=-\left\vert d\sigma^{k}\right\rangle ,
\end{equation}
to compare with,%
\begin{equation}
\left[  \mathbb{B}\right]  \circ\lbrack\mathbb{C}]\circ\left\vert
dx^{k}\right\rangle =\left\vert d\sigma^{k}\right\rangle .
\end{equation}
It follows that,%

\begin{align}
\{\left[  \mathbb{B}\right]  \circ\lbrack\mathbb{C}]+\left[  \Delta\right]
\circ\left[  \mathbb{B}\right]  \}\circ\left\vert dx^{k}\right\rangle  &
=0.\\
\{\left[  \mathbb{B}\right]  \circ\lbrack\mathbb{C}]-\left[  \Delta\right]
\circ\left[  \mathbb{B}\right]  \}\circ\left\vert dx^{k}\right\rangle  &
=2\left\vert d\sigma^{k}\right\rangle \\
\left[  \mathbb{B}\right]  \circ\lbrack\mathbb{C}]  &  =[RH_{1}],\ \\
\left[  \Delta\right]  \circ\left[  \mathbb{B}\right]  \}  &  =\ [LH_{1}].\ \
\end{align}
The remarkable result is that the vector of 1-forms, $\{\left[  \mathbb{B}%
\right]  \circ\lbrack\mathbb{C}]-\left[  \Delta\right]  \circ\left[
\mathbb{B}\right]  \}\symbol{94}\left\vert dx^{k}\right\rangle $, is not
necessarily ZERO, but it is always closed. \ The 1-form can be decomposed into
two component vectors of 1-forms: $[RH_{1}]$ and $[LH_{1}]$. \ It is then
possible that:%

\begin{align}
\{\left[  \mathbb{B}\right]  \circ\lbrack\mathbb{C}]-\left[  \Delta\right]
\circ\left[  \mathbb{B}\right]  \}\symbol{94}\left\vert dx^{k}\right\rangle
&  =2\left\vert d\sigma^{k}\right\rangle \neq0,\\
\{[RH_{1}]-[LH_{1}]\}\symbol{94}\left\vert dx^{k}\right\rangle  &  \neq0,\\
d[RH_{1}]\symbol{94}\left\vert dx^{k}\right\rangle  &  \neq0,\\
d[LH_{1}]\}\symbol{94}\left\vert dx^{k}\right\rangle  &  \neq0,\\
\text{ but }d\{[RH_{1}]-[LH_{1}]\}\symbol{94}\left\vert dx^{k}\right\rangle
&  =2\left\vert dd\sigma^{k}\right\rangle =0.
\end{align}
The exterior differential form, $\{\left[  \mathbb{B}\right]  \circ
\lbrack\mathbb{C}]-\left[  \Delta\right]  \circ\left[  \mathbb{B}\right]
\}\symbol{94}\left\vert dx^{k}\right\rangle $ is a vector of 1-forms that is
differentially closed (has a zero exterior derivative). \ Closed forms
integrated over cycles (not a boundary) lead to values which have integer
ratios. \ They are a form of deRham period integrals, which are related to the
Bohm-Aharanov concepts of quantized circulation.

\subsubsection{Chirality and Topological Quantization}

Perhaps one of the most interesting concepts is that to topological
quantization. \ The idea is that integrals of closed p-forms over closed
domains which are not boundaries have values whose ratios are rational. \ It
is also remarkable that any map to a m dimensional vector valued function can
be used to construct a current a volume element of dimension m.%

\begin{align}
\Psi &  :x^{k}\Rightarrow V^{m}(x^{k}),\\
\Omega_{m}  &  =dV^{1}\symbol{94}...\symbol{94}dV^{m}.
\end{align}
Then it is possible to construct a m-1-form "current", $C$, form the formula:%

\begin{equation}
C=\ i(\rho V^{m})\Omega_{m}%
\end{equation}
If the density, $\rho$ is defined as the inverse of a Holder norm,
$\lambda_{H},$%

\begin{align}
\rho &  =1/\lambda_{H},\\
\lambda_{H}  &  =\{a(V^{1})^{p}+b(V^{2})^{p}+...+\varepsilon(V^{m}%
)^{p}\}^{M/p}%
\end{align}
then the Current has zero divergence for any $p$ and any anisotropic signature
constants, $a,b,...\varepsilon,$%

\begin{align}
J  &  =\ i(V^{m}/\lambda_{H})\Omega_{m},\\
dJ  &  =(div_{m}J)\Omega_{m}=0.
\end{align}
Consider the 2D vector $V=[\varphi,\chi]:$%

\begin{align}
V  &  =[\varphi(x,y,z,T),\chi(x,y,z,T)]\\
\Omega_{2}  &  =d\varphi\symbol{94}d\chi,\\
\lambda_{H}  &  =\{a(\varphi)^{p}+\varepsilon(\chi)^{p}\}^{2/p}\\
J  &  =\{\varphi d\chi-\chi d\phi\}/\lambda_{H}\\
dJ  &  =0.
\end{align}
Note that the conserved current, $J$, consists of two terms,%

\begin{equation}
J1=\varphi d\chi/\lambda_{H},\ \ J2=\chi d\phi/\lambda_{H}\ ,
\end{equation}
neither of which need be closed or exact. \ Yet the difference of the two
terms is always closed. \ This result is another expression of deRham
cohomology theory. \ The divergence condition is valid except in domains where
the Holder norm vanishes. \ The two terms, $J1$ and $J2$, are related to
chirality concepts. \ The method is easily extended to 4D.

A intriguing idea, as yet unexplored, is what do the Holder Norms with
different signatures and exponents, $p$, imply physically. \ There is a hint
that $\varepsilon=-1,$ $p>2$ may be related to diffraction issues.

\subsubsection{The Line Element}

Note that the line element, $\delta s^{2}$, can be constructed from the formula,%

\begin{align}
\delta s^{2}  &  =\left\langle dx^{j}\right\vert \circ\left[  g\right]
\circ\left\vert dx^{k}\right\rangle =\left\langle dx^{k}\right\vert
\circ\lbrack\mathbb{B}]T\circ\lbrack\eta]\circ\left[  \mathbb{B}\right]
\circ\left\vert dx^{k}\right\rangle ,\\
\delta s^{2}  &  =\left\langle \sigma^{m}\right\vert \circ\lbrack\eta
_{mn}]\circ\left\vert \sigma^{n}\right\rangle .
\end{align}
The line element, $\delta s^{2},$ may or may not be integrable. \ When the
infinitesimal mapping generates non-integrable 1-forms, $\left\vert \sigma
^{n}\right\rangle $, the metric $\left[  g\right]  $ is no longer diagonal.
\ The anholonomic terms generated by the infinitessimal mappings lead to
off-diagonal metric coefficients \cite{Vacaru}. \ 

\subsubsection{The Four Forces and the Off-diagonal metric structure.}

A number of years ago, it was noticed that the four forces of physics could be
put into correspondence with the topological structure of the off-diagonal
structure defined in terms of the "timelike" components of the metric,
\cite{rmksubmersive}:%
\begin{equation}
g_{4m}dx^{m}=g_{4x}dx+g_{4y}dx+g_{4z}dz-g_{44}dt.
\end{equation}
The analysis depends upon the topological structure generated by the 1-form,
$g_{4m}dx^{m}$, and its Pfaff Topological Dimension of the off-diagonal
1-form, which can be 1, 2, 3, 4. \ In the original article, the thermodynamic
importance of the differential topological structures were not appreciated.
\ In fact the argument for the PTD\ 1 and 2 cases was made on the basis that
they admitted "infinite range forces", where the PTD\ 3 and 4 cases implied
"short range forces". \ \ Several years later, the concepts of topological
thermodynamics made it apparent that the PTD 1 and 2 cases meant that the
topology was a connected topology, such that infinite range was better stated
as: all points were "reachable". \ In the PTD\ 3 and 4 cases, the
thermodynamic topologies form a disconnected topology. \ Short range was
better stated as: points in a disconnected component of the topology were not
"reachable" from points in another component \ of the disconnected topology.
\ In other words, objects in a disconnected component interact with other
objects in that disconnected component, but not with objects in other
disconnected components. \ Parity is preserved for all but the PTD = 4 case.

The results are displayed below:%

\begin{equation}%
\begin{array}
[c]{ccc}%
\text{\textbf{PTD}} & \text{\textbf{Einstein Solution Examples}} &
\text{\textbf{Thermodynamic system}}\\
1\text{ } & \text{Schwarzschild} & equilibrium\\
2\text{ } & \text{Reissner-Nordstrum} & isolated\\
3\text{ } & \text{Godel} & closed\\
4\text{ } & \text{Kerr-Taub-Nut} & open
\end{array}
\end{equation}%
\begin{equation}%
\begin{array}
[c]{ccc}%
\text{\textbf{Thermodynamic Topology}} & \text{\textbf{Force}} &
\text{\textbf{Parity}}\\
1\text{ long range (connected)} & \text{Gravity} & \text{conserved}\\
2\text{ long range (connected)} & \text{Electromagnetic} & \text{conserved}\\
3\text{ short range (disconnected)} & \text{Strong nuclear } &
\text{conserved}\\
4\text{ short range (disconnected)} & \text{Weak nuclear} & \text{not
conserved}%
\end{array}
\end{equation}

\section{Continuous Topological Evolution}

The usual technique for generating "equations of motion" is to construct a
Lagrange density, $\mathcal{L}$, and integrate it over the volume element,
$\Omega$, then determine "trajectories" that will minimize the integral. \ In
addition, certain constraints can be placed upon the Lagrange density,
producing trajectories that minimize the integral, subject to the constraints.
\ Such constraints lead to the concept of Lagrange multipliers.\ \ 

In this article, another approach is utilized. \ The approach is based upon
the continuous evolution of exterior differential forms (see Vol. 1
\cite{rmklulu} ), and the fact that exterior differential systems contain
topological information. \ The field equations, or equations of motion, must
describe the topological dynamics inherent in the First Law of Thermodynamics.
\ In the next section, the correspondence between the continuous topological
evolution and the First Law is established in terms of the Lie differential
with respect to a process direction field acting on a thermodynamic system
encoded in terms of systems of exterior differential forms.

\subsection{Axioms of Topological Thermodynamics}

The theory of non-equilibrium thermodynamics from the perspective of
continuous topological evolution, as utilized in this monograph, is based on
four axioms.

\begin{quote}
\textbf{Axiom 1}. \ \textit{Thermodynamic physical systems can be encoded in
terms of a 1-form of covariant Action Potentials, }$A_{k}(x,y,z,t...),$%
\textit{\ on a }$\geq$ \textit{four-dimensional abstract variety of ordered
independent variables, }$\{x,y,z,t...\}.$\textit{\ \ The variety supports a
differential volume element }$\Omega_{4}=dx\symbol{94}dy\symbol{94}%
dz\symbol{94}dt...$\textit{\ }\ \medskip

\textbf{Axiom 2.} \ \textit{Thermodynamic processes are assumed to be encoded,
to within a factor, }$\rho(x,y,z,t...)$\textit{, in\ terms of contravariant
vector and/or complex isotropic Spinor direction fields, }$V_{4}%
(x,y,z,t...).\medskip$

\textbf{Axiom 3. }\ \textit{Continuous topological evolution of the
thermodynamic system can be encoded in terms of Cartan's magic formula (see p.
122 in \cite{Marsden}). \ The Lie differential, when applied to an exterior
differential 1-form of Action, \thinspace}$A=A_{k}dx^{k}$\textit{, is
equivalent abstractly to the first law of thermodynamics. }
\end{quote}

\begin{align}
\text{ \textbf{Cartan's Magic Formula} }L_{(\rho\mathbf{V}_{4})}A  &
=i(\rho\mathbf{V}_{4})dA+d(i(\rho\mathbf{V}_{4})A),\\
\text{\textbf{First Law} }  &  :W+dU=Q,\\
\text{\textbf{Inexact Heat 1-form}\ \ }Q  &  =W+dU=L_{(\rho\mathbf{V}_{4}%
)}A,\\
\text{\textbf{Inexact Work 1-form}\ }W  &  =i(\rho\mathbf{V}_{4})dA,\\
\text{\textbf{Internal Energy} \ }U  &  =i(\rho\mathbf{V}_{4})A.
\end{align}
\medskip

\begin{quote}
\textbf{Axiom 4.} \ \textit{Equivalence classes of systems and continuous
processes can be defined in terms of the Pfaff Topological dimension of the
1-forms of Action, }$A$, Work, $W$, and Heat, $Q.$
\end{quote}

In effect, Cartan's methods can be used to formulate precise mathematical
definitions for many thermodynamic concepts in terms of topological properties
- without the use of statistics or geometric constraints such as metric or
connections. \ Moreover, the method applies to non-equilibrium thermodynamical
systems and irreversible processes, again without the use of statistics or
metric constraints. \ The fundamental tool is that of continuous topological
evolution, which is distinct from the usual perspective of continuous
geometric evolution.

\ It is important to remember that thermodynamic processes, encoded by the
symbol, $\mathbf{V}_{4}$, describing a "vector" directional field,\ need not
be characterized by a single parameter (as in Newtonian dynamics). \ In other
words, continuum processes may exhibit topological fluctuations about the
kinematic perfection of "point"\ particles:%

\begin{equation}
\text{Topological fluctuations: }\Delta\mathbf{x=\ }d\mathbf{x}-\mathbf{V}%
_{4}ds\ \mathbf{\neq0.}%
\end{equation}
The concept of a thermodynamic process transcends the concept of diffeomorphic
trajectories, which represent "particle - like" Newtonian curves. \ 

Note that the formula for the First Law is in effect a statement in deRham
cohomology, where the difference between two inexact differential forms is an
exact differential. \ 

\begin{claim}
The Lie differential acting on a thermodynamic system encoded by the system of
forms, $\Xi$, is the dynamical equivalent of the First Law of thermodynamics. \ 
\end{claim}

The processes, $\mathbf{V}_{4}$, can be reversible or irreversible in a
thermodynamic sense. \ The definition of an irreversible process was motivated
by Caratheodory and Morse \cite{Caratheodory}, \cite{Morse}, and is given by
the statement that the heat 1-form, $Q$, does not satisfy the Frobenius
integrability theorem. \ Combining the Caratheodory definition and the Cartan
magic formula yields the expressions:

The necessary condition for thermodynamic irreversibility of a process
$\mathbf{V}_{4}$ acting on a thermodynamic system, $\Xi$, is given by the expression,%

\begin{align}
\text{\textbf{Irreversible Processes}}  &  \text{:}Q\symbol{94}dQ\neq0\\
L_{(\mathbf{V}_{4})}\Xi\symbol{94}L_{(\mathbf{V}_{4})}d\Xi &  =Q\symbol{94}%
dQ\neq0.
\end{align}
\ These requirements can be used to demonstrate that many "dissipative"
systems are not thermodynamically irreversible, but are merely time reversal
invariant. \ Note that thermodynamic processes need not generate a
diffeomorphism. \ In other words, thermodynamic processes can change the
system topology, and that change can be occur continuously.

\subsection{The Thermodynamic Evolution of the Cosmological Vacuum. \ }

Consider the Cosmological Vacuum to be a thermodynamic system encoded by the
infinitesimal mapping equation:%
\begin{align}
\text{The Thermodynamic System}  &  \text{: \ }\left[  \mathbb{B}\right]
\circ\left\vert dx^{k}\right\rangle =\left\vert \sigma^{k}\right\rangle
=e/m_{0}\ \left\vert A^{k}\right\rangle ,\\
\text{Physical dimension of }\sigma^{k}  &  =(\hslash/m_{0}).
\end{align}
The thermodynamic equations of topological evolution are defined in terms of
the action of the Lie differential with respect to of process $\mathbf{V}$
acting on the thermodynamic system:%

\begin{equation}
L_{(\mathbf{V})}\{\left[  \mathbb{B}\right]  \circ\left\vert dx^{k}%
\right\rangle \}=L_{(\mathbf{V})}\{\left\vert \sigma^{k}\right\rangle
\}=e/m_{0}\ L_{(\mathbf{V})}\{\left\vert A^{k}\right\rangle \}
\end{equation}
First, evaluate the bracket factor on the RHS which is assumed to represent
the evolutionary properties of the field intensities:%

\begin{align}
L_{(\mathbf{V})}\{\left\vert A^{k}\right\rangle \}  &  =i(\mathbf{V}%
)\left\vert dA^{k}\right\rangle +d(i(\mathbf{V})\left\vert A^{k}\right\rangle
,\\
&  =i(\mathbf{V})\left\vert F^{k}\right\rangle +d\left\vert U^{k}\right\rangle
=W+dU=Q.
\end{align}
Next, evaluate the LHS which is assume to represent the evolutionary
properties of the field excitations::%
\begin{align}
L_{(\mathbf{V})}\{\left[  \mathbb{B}\right]  \circ\left\vert dx^{k}%
\right\rangle \}  &  =i(\mathbf{V}^{k})d\{\left[  \mathbb{B}\right]
\circ\left\vert dx^{k}\right\rangle \}+d(i(\mathbf{V}^{k})\{\left[
\mathbb{B}\right]  \circ\left\vert dx^{k}\right\rangle \}\\
&  =i(\mathbf{V}^{k})\{\left[  \mathbb{B}\right]  \circ\left[  \mathbb{C}%
\right]  \symbol{94}\left\vert dx^{k}\right\rangle \}+d\{\left[
\mathbb{B}\right]  \circ\left\vert \mathbf{V}^{k}\right\rangle \}\\
&  =\left[  \mathbb{B}\right]  \{i(\mathbf{V}^{k})\left\vert T^{b}%
\right\rangle +(\left[  \mathbb{C}\right]  \circ\left\vert \mathbf{V}%
^{k}\right\rangle +\left\vert d\mathbf{V}^{k}\right\rangle )\}\label{LHS}\\
&  =\left[  \mathbb{B}\right]  \{i(\mathbf{V}^{k})[C^{b}]\symbol{94}\left\vert
dx^{k}\right\rangle -\left[  \mathbb{C}\right]  \circ\left\vert \mathbf{V}%
^{k}\right\rangle \\
&  +\left[  \mathbb{C}\right]  \circ\left\vert \mathbf{V}^{k}\right\rangle
+\left\vert d\mathbf{V}^{k}\right\rangle \}
\end{align}
So the fundamental thermodynamic equation describing the thermodynamic
evolution of the Cosmological Vacuum becomes (a vector equation of 1-forms)
balances the evolution of the field excitations with the evolution with the
field intensities:%
\begin{align}
\text{The Fundamental Thermodynamic Field Equations}  &  \text{:}\\
\text{LHS,\ \ \ \ \ \ \ \ \ \ \ \ \ \ \ \ Excitations }\left[  \mathbb{B}%
\right]  \{i(\mathbf{V}^{k})\left\vert T^{b}\right\rangle +\left\vert
d\mathbf{V}^{k}\right\rangle \}  &  =e/m_{0}\ Q,\\
\text{RHS,\ \ Intensities, where \ }\{i(\mathbf{V}^{k})\left\vert d\sigma
^{k}\right\rangle +d\left\vert i(\mathbf{V}^{k})\sigma^{k}\right\rangle \}  &
=e/m_{0}\ Q.
\end{align}
The three terms in the first bracket of eq. (\ref{LHS}) can be identified in
terms of%

\begin{align}
\text{Affine Torsion "accelerations"}  &  :i(\mathbf{V})\left\vert
T^{b}\right\rangle ,\\
\text{1/2 Coriolis "accelerations"}  &  :\left[  \mathbb{C}\right]
\circ\left\vert \mathbf{V}^{k}\right\rangle ,\\
\,\text{"Acceleration 1-forms" }  &  \text{:}\left\vert Acc^{k}\right\rangle
=\left\vert d\mathbf{V}^{k}\right\rangle
\end{align}
These identifications lead to the fundamental thermodynamic evolution formula
of infinitesimal neighborhoods. \ The "accelerations" of the space-time
properties related to source excitations are balanced by the field intensities
generated by those 1-forms of Action per unit source, that encode the
thermodynamic systems.

It should be remembered that there can be 4 Cartan topologies defined by the
structural equations associated with each of the 4 1-forms that make up the
components of $\left\vert d\sigma^{k}\right\rangle $. \ In examples to be
explored below, certain simplifications can lead to Basis Frames for which
only one of the 4 topological structures possible is not exact. \ The Hopf map
furnishes such an example. \ In addition, there are notable differences
between the Basis Frames that represent the 13 parameter groups of transitive
Particle-Affine maps and intransitive Wave-Affine maps.

\begin{center}
\bigskip

{\Large The Fundamental Formula is of the form:}

.

$\ ${\Large [B]}$\left\vert \text{{\Large Torsion+Coriolis/2+Acc}%
}\right\rangle ${\Large = e/m}$_{\text{0}}${\Large \{i(V)}$\left\vert
\text{{\Large F}}\right\rangle ${\Large +}$\left\vert \text{{\Large dU}%
}\right\rangle ${\Large \} = e/m}$_{\text{0}}${\Large \ Q.\bigskip}
\end{center}

The formula is algebraically correct and should apply to all vector spaces on
space time defined by the Fundamental Axiom. \ It is remarkable that this
thermodynamic equation of evolution (a vector of 1-forms) formally links the
Torsion, Coriolis, and Acceleration 1-forms to the 1-form of heat, $Q,$
generated by the RHS and the 1-form of Action per unit mass (mole),
$\left\vert \sigma^{k}\right\rangle $.

Note that when the Affine Torsion components do not exist, the RHS\ is zero,
and the remaining terms lead to the expression where rotational motion is
balancing an acceleration, in agreement with Newton's laws of circular motion
at constant angular velocity. \ In addition, note that the formula implies
that the concept of Heat requires non-zero valued of the Affine Torsion 2-forms.

\subsection{Processes that Conserve the 4D-Volume Element}

Consider evolutionary processes, $\mathbf{J}$, that conserve the volume element:%

\begin{equation}
L(\mathbf{J})\Omega=di(\mathbf{J})\Omega=0.
\end{equation}
Note that the 3-form defined by the equation,
\begin{equation}
J=i(\mathbf{J})\Omega,
\end{equation}
must be closed if the volume element is to be preserved:%
\begin{equation}
dJ=di(\mathbf{J})\Omega=0.
\end{equation}
Also note that the 3-forms, $\left\vert J^{b}\right\rangle $, defined by the
exterior derivative of the Affine Torsion 2-forms, $\left\vert J^{b}%
\right\rangle =d\left\vert G^{b}\right\rangle $ generates closed 3-forms. \ 

\begin{claim}
Hence the charge current densities generated by the field excitations (the
Affine Torsion 2-forms) preserve the 4D\ volume element. \ 
\end{claim}

\subsubsection{Invariant Volume in Electrodynamic notation}

It is important to recognize that the vector field, $\mathbf{J}_{4}$, has
components directly related to the component functions of the associated
2-form. \ In Electromagnetic notation:%
\begin{align}
i(\mathbf{J}_{4})\Omega &  =\left\vert J^{b}\right\rangle =d\left\vert
G^{b}\right\rangle ,\\
\left\vert \text{div}\mathbf{J}_{4}\Omega\right\rangle  &  =d\left\vert
J^{b}\right\rangle =dd\left\vert G^{b}\right\rangle =0.
\end{align}
Similar results are applicable for the vectors of Topological Torsion,
$\mathbf{T}_{4}$, and Topological Spin,\ $\mathbf{S}_{4}$:%

\begin{align}
\left\vert i(\mathbf{T}_{4})\Omega\right\rangle  &  =\left\vert A\symbol{94}%
dA\right\rangle =\left\vert A\symbol{94}F\right\rangle ,\\
\left\vert \text{div}\mathbf{T}_{4}\Omega\right\rangle  &  =\left\vert
F\symbol{94}F\right\rangle ,\\
\left\vert i(\mathbf{S}_{4})\Omega\right\rangle  &  =\left\vert A\symbol{94}%
G\right\rangle ,\\
\left\vert \text{div}\mathbf{S}_{4}\Omega\right\rangle  &  =\left\vert
F\symbol{94}G\right\rangle -\left\vert A\symbol{94}J\right\rangle .
\end{align}
However, these 4-vectors, $\mathbf{T}_{4}$ and $\mathbf{S}_{4}$, do not
necessarily preserve the 4D-volume element. \ 

\subsubsection{Invariant volume in Hydrodynamic notation}

Similarly, in hydrodynamic notation,%

\begin{align}
i(\mathbf{J}_{4})\Omega &  =\left\vert J^{b}\right\rangle =d\left\vert
\mathfrak{T}^{b}\right\rangle ,\\
\left\vert \text{div}\mathbf{J}_{4}\Omega\right\rangle  &  =d\left\vert
J^{b}\right\rangle =dd\left\vert \mathfrak{T}^{b}\right\rangle =0.
\end{align}
Similar results are applicable for the vectors of Topological Torsion,
$\mathbf{T}_{4}$, and Topological Spin,\ $\mathcal{S}_{4}$:%

\begin{align}
\left\vert i(\mathbf{T}_{4})\Omega\right\rangle  &  =\left\vert \sigma
\symbol{94}d\sigma\right\rangle ,\\
\left\vert \text{div}\mathbf{T}_{4}\Omega\right\rangle  &  =\left\vert
d\sigma\symbol{94}d\sigma\right\rangle ,\\
\left\vert i(\mathcal{S}_{4})\Omega\right\rangle  &  =\left\vert
\sigma\symbol{94}\mathfrak{T}\right\rangle ,\\
\left\vert \text{div}\mathcal{S}_{4}\Omega\right\rangle  &  =\left\vert
d\sigma\symbol{94}\mathfrak{T}\right\rangle -\left\vert \sigma\symbol{94}%
J\right\rangle .
\end{align}
However, these 4-vectors, $\mathfrak{T}_{4}$ and $\mathcal{S}_{4}$, do not
necessarily preserve the 4D-volume element. \ Recall that for the general
problem there are $k$ of these 4-vectors, one for each of 4 Cartan topologies
defined by the four 1-forms, $\left\vert \sigma^{k}\right\rangle .$ \ This
makes symbolic math program like Maple a necessity to overcome the very
tedious algebra. \ Such examples will be presented in part II.

\subsection{Processes that Expand/Contract the 4D Volume element}

\subsubsection{Topological Torsion and conformal invariance of $\Omega$ in
electrodynamic notation}

The concept of Topological Torsion is defined by a Vector
(process),\ $\mathbf{T}_{4}$, and, for the electromagnetic notation, is
composed entirely from the 1-form of Action per unit source, $A$. For each
component of the vector of intensity 1-forms, $\left\vert A^{k}\right\rangle $%

\begin{align}
\left\vert i(\mathbf{T}_{4}^{k})\Omega\right\rangle  &  =\left\vert
A^{k}\symbol{94}F^{k}\right\rangle ,\\
\left\vert \text{div}\mathbf{T}_{4}^{k}\Omega\right\rangle  &  =\left\vert
F^{k}\symbol{94}F^{k}\right\rangle ,
\end{align}
Each of the\ $k$ 4-vectors, $\mathbf{T}_{4}^{k}$, do not necessarily preserve
the 4D-volume element. \ In fact, for each $A^{k}$,%

\begin{align}
L_{(\mathbf{T}_{4})}\Omega &  =\kappa\Omega,\text{ where,}\\
L_{(\mathbf{T}_{4})}\Omega &  =\{2(\mathbf{B\circ E)}\}_{electromagnetism}%
\Omega
\end{align}

\subsubsection{Topological Torsion and conformal invariance of $\Omega$ in
hydrodynamic notation}

The concept of Topological Torsion in hydrodynamic notation is composed
entirely from the 1-form of Action per unit source, $\sigma$. For each
component of the vector of intensity 1-forms, $\left\vert \sigma
^{k}\right\rangle $%

\begin{align}
\left\vert i(\mathbf{T}_{4}^{k})\Omega\right\rangle  &  =\left\vert \sigma
^{k}\symbol{94}d\sigma^{k}\right\rangle ,\\
\left\vert \text{div}\mathbf{T}_{4}^{k}\Omega\right\rangle  &  =\left\vert
d\sigma^{k}\symbol{94}d\sigma^{k}\right\rangle ,
\end{align}
Each of the\ $k$ 4-vectors, $\mathbf{T}_{4}^{k}$, do not necessarily preserve
the 4D-volume element. \ In fact, for each $\sigma^{k}$,%

\begin{align}
L_{(\mathbf{T}_{4})}\Omega &  =\kappa\Omega,\text{ where,}\\
L_{(\mathbf{T}_{4})}\Omega &  =\{2(vorticity\circ
acceleration)\}_{hydrodynamics}\Omega\text{.}%
\end{align}
When the coefficient, $\kappa$\thinspace, is not zero, the processes acting on
the volume element is said to produce conformal invariance. \ When $\kappa$ is
a constant, it can be interpreted as a homogeneity index of degree $\kappa$.
\ Note that $\kappa$ need not be an integer, which can then be interpreted as
a fractal self similarity condition. \ Further note that $\kappa$ can be both
spatially and time dependent, ultimately leading to zero value, and stability
of the volume element.

\begin{claim}
This result indicates that a topological torsion process, $\mathbf{T}_{4}^{k}%
$, can cause the cosmological vacuum to expand (or contract) irreversibly,
depending on the sign of the bulk viscosity coefficient, $2(\mathbf{B\circ
E)\approx2(\omega\circ a)}$
\end{claim}

The contraction or expansion of the Universe (the 4D volume element) has been
expressed in terms of a "dilaton" field, but the concept has its engineering
roots in terms of "bulk viscosity" \cite{Eckart}, with out reference to
General Relativity. \ The engineering idea forces attention on the fact that
dilaton methods in General Relativity are thermodynamically irreversible. \ 

Recall that for the general problem there are $k$ of these 4-vectors, one for
each of 4 Cartan topologies defined by the four 1-forms, $\left\vert
\sigma^{k}\right\rangle .$ \ This makes symbolic math program like Maple a
necessity to overcome the very tedious algebra. \ A number of Maple examples
will be attached as Part II.

\subsubsection{Topological Spin and conformal invariance of $\Omega$ in
electrodynamic notation}

The concept of Topological Spin is defined by a Vector (process),\ $\mathbf{S}%
_{4}$, that is composed from both the components of the 1-forms, $\left\vert
A^{k}\right\rangle $ and the vector of two forms representing Affine Torsion,
$\left\vert G^{k}\right\rangle $. \ Results similar to those enumerated above
are applicable for the vectors of Topological Torsion, $\mathbf{S}_{4}$. \ For
each component of the vector of intensity 1-forms, $\left\vert A^{k}%
\right\rangle $%

\begin{align}
\left\vert i(\mathbf{S}_{4}^{k})\Omega\right\rangle  &  =\left\vert
A^{k}\symbol{94}G^{k}\right\rangle ,\\
\left\vert \text{div}\mathbf{S}_{4}^{k}\Omega\right\rangle  &  =\left\vert
F^{k}\symbol{94}G^{k}\right\rangle -\left\vert A^{k}\symbol{94}J^{k}%
\right\rangle ,\\
\left\vert J^{k}\right\rangle  &  =d\left\vert G^{k}\right\rangle .
\end{align}
In electromagnetic notation, the 4-divergence of the Topological Spin process
is to be recognized as the Field Lagrange density\footnote{Note that
expression is not $\{\mathbf{B\circ H-D\circ E\}/2}$ (see \cite{rmkintrinsic}%
).}, minus the Field (charge current and potential) interactions.
\begin{equation}
\left\vert \text{div}\mathbf{S}_{4}^{k}\Omega\right\rangle =(\{\mathbf{B\circ
H-D\circ E\}-\{A\circ J-}\rho\phi\})\Omega
\end{equation}

\begin{claim}
This result indicates that a topological Spin process, \textbf{$T$}$_{4}$, can
cause the cosmological vacuum to expand (or contract), depending on the sign
of the Lagrange coefficient, $\{\mathbf{B\circ H-D\circ E\}-\{A\circ J-}%
\rho\phi\}$.
\end{claim}

It is obvious that a combination of the two processes should be studied. \ %

\begin{align}
\mathbf{E}_{4}  &  =\mathbf{T}_{4}+(\hslash/e^{2})\mathbf{S}_{4}.\\
L_{(\mathbf{E}_{4})}\Omega &  =\{2(\mathbf{B\circ E)+}(\hslash/e^{2}%
)(\mathbf{B\circ H-D\circ E-A\circ J+}\rho\phi)\}\Omega.
\end{align}
Note that coefficient $(\hslash/e^{2})$ is the Hall coefficient, and is used
for dimensional equivalence in the composite formula. \ Note that the
identification of Affine Torsion with the excitation field, $\left\vert
G^{k}\right\rangle $, permits the formalism presented to be interpreted in
terms of either topological electromagnetism, or topological hydrodynamics. \ 

\begin{claim}
In classical hydrodynamics, the concept of the excitation fields has "slipped
through the net". \ Herein it is recognized that the concept of Affine Torsion
defines the excitation fields for both hydrodynamics and electromagnetism.
\ Hence the concept of Topological Spin can be developed for classical
hydrodynamic systems, when Affine Torsion is taken into account. \ However,
Affine Torsion 2-forms are necessary, but not sufficient, for the creation of
3-forms of Topological Spin.
\end{claim}

It is remarkable that a combination of the Topological Spin and the
Topological Torsion processes could influence the expansion rate of the
universe. \ In particular it is conceivable that the two process paths, each
of which is not divergence free, could yield a composite that is divergence
free. \ This concept is another exhibition that there can exist two non-exact
forms with a difference that is exact. \ Simply said, the idea is that a
"rotation" can balance a "contraction". \ Resaid, the divergence of
Topological Spin - a rotation due to a non-zero Lagrangian - could balance
divergence of \ Topological Torsion - a contraction due to Bulk viscosity.

\section{Diffusion and Conformal Evolution}

In this section attention will be paid to the RHS of the fundamental equation,
where certain constraints will produce a form of a diffusion equation. \ This
idea goes back to Bateman \cite{Bateman}, where he demonstrated that solutions
to a diffusion equation in 2+1 space-time could be transformed into solutions
to the wave equation in 3+1 space-time. \ It is easier (for me)\ to use the
topological formulation of electromagnetism (which is formally equivalent to
the topological theory of hydrodynamics). \ In thermodynamics the fundamental
unit source is the mole. \ In electromagnetism the fundamental unit source is
the electric charge; \ division of the 1-form of Action, $\left\vert
A^{k}\right\rangle $, by the factor $(\hslash/e)$ leads to equations that are
free from "physical dimensions" \ In hydrodynamics, the fundamental unit
source is mass, and the division of the 1-form of Action, $\left\vert
\sigma^{k}\right\rangle $, by the "kinematic viscosity", $(\hslash/m)$, leads
to a Reynolds number type of formulation, where the Action 1-form is
dimensionless in terms of the "physical dimensions". \ 

To reduce the algebra, only one component of the vector of field intensities,
$\left\vert A^{t}\right\rangle $, will be treated as a non-exact 1-form of
Pfaff Topological Dimension 4. \ The Basis Frame will be a member of the
W-Affine Group. \ The thermodynamic evolution of the field potentials
generated by $A^{t}$ are given by the first law as:%

\begin{equation}
L(\mathbf{V}_{4})A^{t}=i(\mathbf{V}_{4})dA^{t}+d(i(\mathbf{V}_{4})A^{t}=Q.
\end{equation}
The thermodynamic evolution of the limit points $dA^{t}$ are given by the expression,%

\begin{equation}
L(\mathbf{V}_{4})dA^{t}=d(i(\mathbf{V}_{4})dA^{t})=dQ.
\end{equation}

The next step is to reconsider those processes which preserve the 4D\ volume
element, and those which permit an expansion (or contraction) of the
4D\ volume element in a homogenous manner. \ There are three distinct classes
of processes, each of which can be associated with a minimal hypersurface
(mean curvature = 0). \ The wave equation, diffusion equation, and the
equation of a minimal surface are all related to a form of a null divergence
condition on a vector field, and are thereby related to some form of a
conservation law, or minimization process. For example consider the variety
$\{x,y,z;\xi)$ with a volume N-form, $\Omega=dx\symbol{94}dy\symbol{94}%
dz\symbol{94}d\xi$. Also consider a contravariant vector field $J$ with
components
\begin{equation}
J=\rho V=\rho(x,y,z,t)[V^{x},V^{y},V^{z},1]
\end{equation}
Then the volume $\Omega$ is an invariant with respect to an evolutionary path
generated by $J=\rho V$ if $div_{4}J=0$. That is,%

\begin{align}
L(\rho V)\Omega &  =d(i(J)\Omega\}=\{div_{4}J\}\Omega\\
&  =\{\partial(\rho V^{x})/\partial x,\partial(\rho V^{y})/\partial
y,\partial(\rho V^{3})/\partial z,\partial\rho/\xi\}\Omega\\
&  \Rightarrow0,\text{when \ }\{div_{4}J\}=0.
\end{align}
Therefor, the invariant volume element, $\Omega,$ is associated with a
\ process that has zero 4-divergence. \ 

However, suppose the process does not have zero divergence, but is equal to
some function, $\Psi$. \ Then, the differential volume element is not
invariant with respect to the process, but it is conformal:%

\begin{equation}
L(\rho V)\Omega=d(i(J)\Omega\}=\{div_{4}J\}\Omega=\Psi\Omega.
\end{equation}

\subsection{The Wave Equation}

Examine the case where:
\begin{equation}
\mathbf{V}_{4}=\rho\lbrack\partial\phi/\partial x^{k};\varepsilon
S]\ \ \ \rho=constant,\ \ \ S=\partial\phi/\partial\xi\text{.}%
\end{equation}
Then the null divergence condition becomes:%
\begin{equation}
div_{4}J=\rho~div_{4}V=\rho\{\partial^{2}\phi/\partial x^{2}+\partial^{2}%
\phi/\partial y^{2}+\partial^{2}\phi/\partial z^{2}+\varepsilon\partial
^{2}\phi/\partial\xi^{2}\}\Rightarrow0.
\end{equation}
When $\varepsilon=-1$ and $\xi=ct$, the null divergence constraint is exactly
the wave equation.

Suppose the density function was not constant. Then the zero divergence
condition would be given by the expression:%

\begin{equation}
div_{4}J=div_{4}(\rho V)=\{\partial^{2}\phi/\partial x^{2}+\partial^{2}%
\phi/\partial y^{2}+\partial^{2}\phi/\partial z^{2}+\varepsilon\partial
^{2}\phi/\partial\xi^{2}\}+(grad_{4}\ln\rho)\circ V=0.
\end{equation}
This equation describes a modified wave equation, but if the density function
is a first integral (a process invariant) then the term $(grad_{4}\ln
\rho)\circ V$ vanishes, and the standard form of the wave equation is recovered.

\subsection{The Diffusion Equation}

Similarly, examine the case\footnote{k =1,2,3} where:
\begin{equation}
\mathbf{V}_{4}=\rho\lbrack\partial\psi/\partial x^{k};\varepsilon
\psi]\ \ \ \rho=constant,\ \ \ S=\phi\text{.}%
\end{equation}
\ then the null divergence condition becomes:%

\begin{equation}
div_{4}J=\rho~div_{4}V=\rho\{\partial^{2}\psi/\partial x^{2}+\partial^{2}%
\psi/\partial y^{2}+\partial^{2}\psi/\partial z^{2}+\varepsilon\partial
\psi/\partial\xi\}\Rightarrow0.
\end{equation}
When $\varepsilon=-1$ and $\xi=t/D$, the null divergence constraint is exactly
the diffusion equation,%
\begin{equation}
D\partial\psi/\partial t=\partial^{2}\psi/\partial x^{2}+\partial^{2}%
\psi/\partial y^{2}+\partial^{2}\psi/\partial z^{2}%
\end{equation}
Again if the density is not a constant, then a modified diffusion equation is
the result,%
\begin{equation}
\{\partial^{2}\psi/\partial x^{2}+\partial^{2}\psi/\partial y^{2}+\partial
^{2}\psi/\partial z^{2}+\varepsilon D\partial\psi/\partial t\}+(grad_{4}%
\ln\rho)\circ V\Rightarrow0.
\end{equation}

\subsection{The Minimal Surface equation}

Examine the case where ($\lambda_{H}$ is a Holder norm) relative to any vector
field, $\mathbf{V}_{4}$ :
\begin{align}
\mathbf{V}_{4}  &  =[V^{k};V^{s}],\ \ \ \ ,\ \ \ \ \rho\mathbf{V}%
_{4}=\mathbf{V}_{4}/\lambda_{H},\ \ \\
\lambda_{H}  &  =[a(V^{x})^{p}+b(V^{y})^{p}+c(V^{z})^{p}+\varepsilon
(V^{s})^{p}]^{N/p}.
\end{align}
Consider the 3-form constructed from the equation,%
\begin{equation}
C=i(\mathbf{V}_{4}/\lambda_{H})dV^{x}\symbol{94}dV^{y}\symbol{94}%
dV^{z}\symbol{94}dV^{s}=i(\mathbf{V}_{4}/\lambda_{H})\Omega_{(\mathbf{V}_{4}%
)}.
\end{equation}
For any anisotropic signature (a,b,c,$\epsilon$) and any exponent p, then if N
= 4, the 3-form is homogeneous of degree zero. \ It follows that the trace of
the Jacobian matrix $[Jac(\mathbf{V}_{4}/\lambda_{H})]$ is zero. \ The surface
defined by the characteristic polynomial of $[Jac(\mathbf{V}_{4}/\lambda
_{H})]$ is a 4th order polynomial = 0, representing a hypersurface in 4D,
which has zero mean curvature. \ A hypersurface with zero mean curvature is a
Minimal Surface. \ Hence the renormalized vector, $(\mathbf{V}_{4}/\lambda
_{H})$, has zero divergence on $\Omega_{(\mathbf{V}_{4})}$ for all values of
the Holder norm, if N = 4. \ 

It has been assumed that the determinant of $[Jac(V/\lambda_{H})]$ is maximal
rank and non-zero. \ Therefor, the zero divergence condition is also valid
on\ differential volume element, $\Omega=dx\symbol{94}dy\symbol{94}%
dz\symbol{94}ds.$ \ The wave equation and the diffusion equation are special
cases of the minimal surface equation. \ 

\subsection{The Characteristic Polynomial}

In 4D, the characteristic polynomial of $[Jac(\mathbf{V}_{4}/\lambda_{H})]$ is
of 4th degree, where for (possibly complex) eigen values, $\gamma$, the
polynomial (by the Cayley Hamilton theorem) generates the hypersurface,%

\begin{align}
Ch[Jac(\mathbf{V}_{4}/\lambda_{H})]  &  =\gamma^{4}-M\gamma^{3}+G\gamma
^{2}-A\gamma+K=0,\\
&  =\gamma^{4}-(4-N)\gamma^{3}/\lambda_{H}+(6-3N)\gamma^{2}/\lambda_{H}%
^{2}-(4-3N)\gamma/\lambda_{H}^{3}+(1-N)/\lambda_{H}^{4}.\\
&  =(\gamma\lambda_{H}-1)^{3}\left(  \gamma\lambda_{H}-1+N\right)  =0.
\end{align}
For different values of N, the Holder norm creates different homogeneity
criteria. \ It was demonstrated above that for a minimal surface the trace of
the matrix $\,[Jac(\mathbf{V}_{4}/\lambda_{H})]$ vanishes. \ It is easily
demonstrated (with Maple) that%

\begin{align}
\text{Mean Curvature}\text{: }  &  M=(4-N)/\lambda_{H}=Trace\ [Jac(\mathbf{V}%
_{4}/\lambda_{H})],\\
\text{Gauss Curvature}\text{: }  &  G=(6-3N)/\lambda_{H}^{2},\\
\text{Cubic Curvature}\text{: }  &  A=(4-3N)/\lambda_{H}^{3},\\
\text{ Quartic Curvature}\text{: \ }  &  K=(1-N)/\lambda_{H}^{4},
\end{align}
So, as mentioned above, for N = 4, the trace of $Ch[Jac(\mathbf{V}_{4}%
/\lambda_{H})$ vanishes and the Mean Curvature is zero of the hypersurface is
zero; \ the 4-divergence of the process $\mathbf{V}_{4}/\lambda_{H}$ is zero,
\ and the volume element is invariant for such all such homogeneous processes (N=4).

If the Mean curvature is not zero, then divergence of the homogeneous process
depends upon the Holder norm, $\lambda_{H}$. \ Hence the conformal factor,
$\Psi$, in the equation,
\begin{align}
L(\rho V)\Omega &  =d(i(J)\Omega\}=\{div_{4}J\}\Omega=\Psi\Omega,\\
\{div_{4}J\}  &  =\Psi=(4-N)/\lambda_{H}\neq0.
\end{align}
is well defined for any N, and for all forms of the Holder norm. \ When N
%TCIMACRO{\TEXTsymbol{>} }%
%BeginExpansion
$>$
%EndExpansion
4 the volume is contracting; \ when N
%TCIMACRO{\TEXTsymbol{<} }%
%BeginExpansion
$<$
%EndExpansion
4, the volume is expanding, due to the homogeneous process $\mathbf{V}%
_{4}/\lambda_{H}$.

Now return to the "diffusion" format,
\begin{align}
\mathbf{V}_{4}  &  =[\partial\psi/\partial x,\partial\psi/\partial
y,\partial\psi/\partial z,\varepsilon\psi],\ \ \ \ \rho\mathbf{V}%
_{4}=\mathbf{V}_{4}/\lambda_{H}\ \ ,\\
\lambda_{H}  &  =[a(\partial\psi/\partial x)^{p}+b(\partial\psi/\partial
y)^{p}+c(\partial\psi/\partial z)^{p}+\varepsilon(\psi)^{p}]^{N/p}.
\end{align}
Then the divergence of $\mathbf{V}_{4}/\lambda_{H}$ then yields a modified
diffusion equation:%

\begin{align}
div_{4}(\mathbf{V}_{4}/\lambda_{H})  &  =(div_{4}\mathbf{V}_{4})/\lambda
_{H}-\mathbf{V}_{4}\circ grad(\lambda_{H})/(\lambda_{H})^{2}=(4-N)/\lambda
_{H},\\
&  =(div_{4}\mathbf{V}_{4})-\mathbf{V}_{4}\circ grad_{4}(\lambda_{H}%
)/(\lambda_{H})=(4-N)/\lambda_{H},\\
-(\varepsilon/D)\partial\psi/\partial t  &  =\partial^{2}\psi/\partial
x^{2}+\partial^{2}\psi/\partial y^{2}+\partial^{2}\psi/\partial z^{2}%
-(4-N)-\mathbf{V}_{4}\circ grad_{4}(\ln\rho).
\end{align}
\ 

If the Characteristic Polynomial defines a minimal surface $M=0$ for $N=4$,
then the Gauss curvature is negative for any signature, indicating the surface
is unstable. \ The volume element is invariant, but the Minimal Surface is
unstable. \ Suppose that $N\leq2$; then the Gauss curvature is positive,
indicating that the Hypersurface generated by the Characteristic polynomial is
stable, \ but it is not a Minimal surface. \ However, the Volume element is
expanding. \ 

\begin{conjecture}
Is the Expansion of the universe required to stabilize the Hypersurface
generated by the Characteristic polynomial?
\end{conjecture}

There is one singular case where the homogeneity index N\ = 0. \ Then the 4D
differential volume element becomes singular. \ However, the analysis can be
extended to the 3D case:{}%

\begin{align}
\mathbf{V}_{3}  &  =[U,V,W],\\
\lambda_{H}  &  =[a(U)^{p}+b(V)^{p}+c(W)^{p}]^{n/p}\\
Ch\,[Jac(\mathbf{V}_{3}/\lambda_{H})  &  =\gamma^{3}-M\gamma^{2}+G\gamma
^{1}-A=0.\\
\text{Mean Curvature}  &  \text{: }M=(3-n)/\lambda_{H},\\
\text{Gauss Curvature}  &  \text{: }G=(3-2n)/\lambda_{H}^{2},\\
\text{Cubic Curvature}  &  \text{: }A=(1-n)/\lambda_{H}^{3},
\end{align}
where the curvature similarity invariants are independent from the constant
anisotropy coefficients, $\{a,b,c\}$ and the exponent $p$. \ 

\section{The Continuum Field for a Plasma (or a Fluid)}

\subsection{Affine Torsion and Excitation 2-forms}

As mentioned above, the vector of Affine Torsion 2-forms $\left\vert
\mathfrak{T}^{b}\right\rangle =\left[  \mathbb{C}_{a}^{b}\right]
\symbol{94}\left\vert dx^{a}\right\rangle \ $has coefficients that can be put
into 1-1 correspondence (to within a factor) with the concept of "excitation
fields" $\left\vert G^{b}(D,H)\right\rangle $ in classical electromagnetism.
\ \ It is important to realize that the use of the words "Affine torsion" to
describe the antisymmetric coefficients of a Cartan connection is unfortunate,
and has nothing to do with whether or not the Basis Frame matrix is a member
of the Affine group, or one of its subgroups. \ Classically, the affine group
is a \textit{transitive} group of 13 parameters in 4D, (see p.162 in Turnbull
\cite{Turnbull}). \ The anti-symmetry concept related to Affine Torsion is
described by the same formula that defines the vector of excitation 2-forms,
$\left\vert \mathfrak{T}^{b}\right\rangle =\left[  \mathbb{C}_{a}^{b}\right]
\symbol{94}\left\vert dx^{a}\right\rangle ,$ for any Basis Frame $\left[
\mathbb{B}(x)\right]  $. \ \ For example, the torsion\ formula holds equally
well for Basis Frames which are elements of the 15 parameter projective group,
which is not affine.
\begin{align}
\text{{\small (Affine) \ }Torsion 2-forms\ \ \ }\left\vert \mathfrak{T}%
^{b}\right\rangle  &  =\left[  \mathbb{C}\right]  \symbol{94}\left\vert
dx^{a}\right\rangle ,\\
\text{Vector of Field Excitation 2-forms \ }  &  =\left\vert \mathfrak{T}%
^{b}\right\rangle \\
&  =\left[  \mathbb{B}\right]  ^{-1}\circ\left\vert d\sigma^{k}\right\rangle ,
\end{align}
The vector of 2-forms $\left\vert \mathfrak{T}^{b}\right\rangle $ is formally
equivalent (for each index $b$)\ to the (impair, or odd) 2-form (density) of
the field excitations (D and H) in electromagnetic theory (see Vol. 4
\cite{rmklulu}). \ In the notation of electromagnetism, the source of field
excitations (and, consequently, topological charge and and topological spin)
is due to the Affine Torsion components of the Cartan Connection. \ 

Note that the matrix $\left[  \mathbb{B}\right]  ^{-1}$ plays the role of the
Constitutive map between $\mathbf{E,B}$ and $\mathbf{D,H}$. \
\begin{align}
\left\vert \mathfrak{T}^{b}\right\rangle  &  =\left[  \mathbb{B}\right]
^{-1}\circ\left\vert d\sigma^{k}\right\rangle ,\label{CM}\\
\left[  \mathbb{B}\right]  ^{-1}  &  \approx\text{a Constitutive map}%
\end{align}
If the global (integrability) assumption, \ $\left[  \mathbb{F}(x)\right]
\circ\left\vert x^{a}\right\rangle \ \Rightarrow\ \left\vert y^{k}%
\right\rangle ,$ is imposed, then it is possible by exterior differentiation
to show that a constraint must be established between the excitation 2-forms
and the Cartan Curvature 2-forms constructed from the globally integrable
Basis Frames, $\left[  \mathbb{F}(x)\right]  :$%

\begin{align}
\left[  \mathbb{F}(x)\right]  \circ\{[\mathbb{C}_{\mathbb{F}}]\circ\left\vert
x^{a}\right\rangle \ +\left\vert dx^{a}\right\rangle \}  &  =\ \left\vert
dy^{k}\right\rangle ,\\
\text{ such that \ }[\mathbb{C}_{\mathbb{F}}]\circ\left\vert dx^{a}%
\right\rangle  &  =-\{d[\mathbb{C}_{\mathbb{F}}]+[\mathbb{C}_{\mathbb{F}%
}]\symbol{94}[\mathbb{C}_{\mathbb{F}}]\}\circ\left\vert x^{a}\right\rangle \ .
\end{align}
Hence as the vector of excitation 2-forms $\left\vert G^{b}\right\rangle $ has
been defined in terms of the Cartan Connection, two different results are
obtained for the two different types of Basis Frames:
\begin{align}
\left\vert \mathfrak{T}^{b}\right\rangle _{\mathbb{B}}  &  =[\mathbb{C}%
_{\mathbb{B}}]\symbol{94}\left\vert dx^{a}\right\rangle \\
\left\vert \mathfrak{T}^{b}\right\rangle _{\mathbb{F}}  &  =[\mathbb{C}%
_{\mathbb{F}}]\symbol{94}\left\vert dx^{a}\right\rangle .
\end{align}
The integrability condition places a constraint on the Cartan Curvature and
the Affine torsion coefficients of the Cartan Connection, $[\mathbb{C}%
_{\mathbb{F}}]$, which is not equivalent to the constitutive map (eq \ref{CM})
given above:%

\begin{align}
\left\vert \mathfrak{T}^{b}\right\rangle _{\mathbb{F}}  &  =-\{d[\mathbb{C}%
]+[\mathbb{C}]\symbol{94}[\mathbb{C}]\}\left\vert x^{a}\right\rangle \\
&  \neq\left[  \mathbb{B}\right]  ^{-1}\circ\left\vert d\sigma^{k}%
\right\rangle .
\end{align}
The result demonstrates that the set of infinitesimal Basis Frames is much
different from the global set of Basis Frames.

None of this development depends upon the explicit specification of a metric,
reinforcing the fact that Maxwell's theory of Electrodynamics is a
topological, not a geometric theory. \ Again, remember that the
electromagnetic notation is used as a learning crutch to emphasize the
universal ideas of the Cosmological Vacuum. \ The formulas are valid
topological descriptions of the field structures of all continuum "fluids". \ 

Herein, for simplicity, it is assumed that all functions of the Basis Frame
are at least C2. \ However, note that the definitions of the matrix of
\ connection 1-forms $[\mathbb{C}]$ and the vector of 2-forms $\left\vert
d\sigma^{k}\right\rangle $ only require C1 functions.

\subsection{The Lorentz Force and the Lie differential (EM notation)}

The Lorentz force is a \textit{derived, universal,} concept in terms of the
thermodynamic cohomology. \ It is generated by application of Cartan's Magic
formula \cite{Marsden} to the 1-form $A$ that that encodes all or part of a
thermodynamic system. \ The system, $A$, can be interpreted as an
electromagnetic system, or a hydrodynamic system, or any other system that
supports continuous topological evolution. \ For the purposes herein apply
Cartan's magic formula to the formula for infinitesimal mapping produced by
the matrix multiplication of a vector of perfect (exact) differentials of the
base variables given by eq(\ref{diffmap}).%

\begin{align}
\left[  \mathbb{B}_{a}^{k}(y)\right]  \circ\left\vert dy_{k}\right\rangle \
&  \Rightarrow\ \left\vert A^{k}(y,dy)\right\rangle \\
L_{(V)}\{\left[  \mathbb{B}_{a}^{k}(y)\right]  \circ\left\vert dy_{k}%
\right\rangle \}  &  =L_{(V)}\left\vert A^{k}(y,dy)\right\rangle .
\end{align}

Recall that the exterior derivative of any specific 1-form, $A^{k}$, if not
zero, can be defined as a 2-form with coefficients of the type%

\begin{align}
F  &  =dA=\{\partial A_{k}/\partial x^{j}-\partial A_{j}/\partial
x^{k}\}dx^{j}\symbol{94}dx^{k}=F_{jk}dx^{j}\symbol{94}dx^{k}\\
&  =\mathbf{B}_{z}dx\symbol{94}dy+\mathbf{B}_{x}dy\symbol{94}dz+\mathbf{B}%
_{y}dz\symbol{94}dx+\mathbf{E}_{x}dx\symbol{94}dT+\mathbf{E}_{y}%
dy\symbol{94}dT+\mathbf{E}_{z}dz\symbol{94}dT.\nonumber
\end{align}
The specialized notation for the coefficients used above is that\ often used
in studies of electromagnetism, but the topological 2-form concepts are
universal, independent from the notation. \ 

Given any process that can be expressed in terms of a vector direction field,
$V=\rho\lbrack\mathbf{V},1]$, and for a physical system, or component of a
physical system, that can be encoded in terms of a 1-form of Action, $A,$ the
topological evolution of the 1-form relative to the direction field can be
described in terms the Lie differential:%
\begin{align}
L_{(V)}A  &  =i(V)dA+d(i(V)A)\\
&  =i(V)F+d(i(V)A)\\
&  =\rho\{\mathbf{E}+\mathbf{V}\times\mathbf{B)}_{k}dx^{k}-\rho\{\mathbf{E}%
\cdot\mathbf{V}\}dT\\
&  +d(\rho\mathbf{A\cdot V}-\rho\phi)\\
&  =\text{Work due to Lorentz force - dissipative power }\\
&  \text{+ change of internal interaction energy}.
\end{align}
Note that if the notation is changed (such that the vector potential is
designated as the velocity components of a fluid), then the "Lorentz force"
represents the classic expression to be found in the\ formulation of the
hydrodynamic Lagrange Euler equations of a fluid (see Vol. 3 \cite{rmklulu}).
\ A fluid, based upon a 1-form of Action of Pfaff Topological dimension 2 (or
greater) obeys the topological equivalent of a Maxwell Faraday induction law!

\begin{remark}
The\ universal concept of a Lorentz force is derived from the properties of a
"Cosmological Vacuum", and does not require a separate postulate of existence.
\end{remark}

\subsection{Processes that leave the intensity 2-form, $F\,$, invariant, or
conformally invariant}

\bigskip If a process $\mathbf{V}_{4}$ is to preserve the 2-form of field
intensities, $F$, then%

\begin{equation}
L_{(\mathbf{V}_{4})}dA=L_{(\mathbf{V}_{4})}F=d(i(\mathbf{V}_{4})dA)=dQ=0.
\end{equation}
This constraint identifies the process as expressing the Helmholtz
theorem\ (conservation of vorticity) in hydrodynamics. \ 

Now consider the equations in EM format,
\begin{align}
\mathbf{V}_{4}  &  =[\mathbf{V}_{3},1]\\
i(\mathbf{V}_{4})F  &  =-\{(\mathbf{E+V}_{3}\mathbf{\times B})_{k}%
dx^{k}-(\mathbf{E\circ V}_{3})dT\},\\
di(\mathbf{V}_{4})F  &  =(\partial(\mathbf{E+V}_{3}\mathbf{\times B})/\partial
T+\nabla(\mathbf{E\circ V}_{3}))_{k}dT\symbol{94}dx^{k}+\\
&  \{curl(\mathbf{E+V}_{3}\mathbf{\times B})\}^{x}dy\symbol{94}dz\\
&  +\{curl(\mathbf{E+V}_{3}\mathbf{\times B})\}^{y}dz\symbol{94}dx\\
&  +\{curl(\mathbf{E+V}_{3}\mathbf{\times B})\}^{z}dx\symbol{94}dy\text{,
where}\\
(curl(\mathbf{E+V}_{3}\mathbf{\times B})  &  =-\partial\mathbf{B}/\partial
T+\mathbf{B\circ\lbrack}\nabla\mathbf{V}_{3}]-\mathbf{V}_{3}\circ\lbrack
\nabla\mathbf{B]}+(\text{div}\mathbf{V}_{3})\mathbf{B}%
\end{align}
If $dQ$ is equal to zero, then there are 6 equations of constraint:
\begin{align}
\partial(\mathbf{E+V}_{3}\mathbf{\times B})/\partial T+\nabla(\mathbf{E\circ
V}_{3})  &  =0,\\
-\partial\mathbf{B}/\partial T+\mathbf{B\circ\lbrack}\nabla\mathbf{V}%
_{3}]-\mathbf{V}_{3}\circ\lbrack\nabla\mathbf{B]}+(\text{div}\mathbf{V}%
_{3})\mathbf{B}  &  =0.
\end{align}
It is apparent that the last equation represents a diffusion equation for the
components of the vector $\mathbf{B}$. \ This method was utilized in a 2+1
space to demonstrate that the Schroedinger equation admitted and exact mapping
that permitted the square of the wave function to be interpreted as the
Enstrophy (square of the vorticity) in a hydrodynamic format \cite{rmkBohm}.
\ The method is remindful of Ricci flows, but does not depend upon metric explicitly.

The next concept is to examine conformal invariance of the 2-form, $F$. \ The
equations are similar to those above:%

\begin{equation}
L_{(\mathbf{V}_{4})}dA=L_{(\mathbf{V}_{4})}F=d(i(\mathbf{V}_{4})dA)=\varkappa
F.
\end{equation}
The analysis leads to the constraints:%

\begin{align}
\partial(\mathbf{E+V}_{3}\mathbf{\times B})/\partial T+\nabla(\mathbf{E\circ
V}_{3})  &  =\varkappa\mathbf{E},\\
-\partial\mathbf{B}/\partial T+\mathbf{B\circ\lbrack}\nabla\mathbf{V}%
_{3}]-\mathbf{V}_{3}\circ\lbrack\nabla\mathbf{B]}+(\text{div}\mathbf{V}%
_{3})\mathbf{B}  &  =\varkappa\mathbf{B}.
\end{align}
Again, the conformality factor can represent a homogeneity condition, or a
fractal self-similarity condition under the evolutionary process. \ Such
processes are modifications of a diffusion equation.

\subsection{Processes that leave the excitation 2-form, $G$, invariant, or
conformally invariant}

If a process $\mathbf{V}_{4}$ is to preserve the 2-form of field excitations,
$G$, then%

\begin{equation}
L_{(\mathbf{V}_{4})}G=i(\mathbf{V}_{4})dG+d(i(\mathbf{V}_{4})G=0.
\end{equation}
This constraint identifies the process as expressing the Helmholtz
theorem\ (conservation of vorticity) in hydrodynamics. \ 

Now consider the equations in EM format,
\begin{align}
\mathbf{V}_{4}  &  =[\mathbf{V}_{3},1]\\
i(\mathbf{V}_{4})G  &  =-\{(\mathbf{H+V}_{3}\mathbf{\times D})_{k}%
dx^{k}-(\mathbf{H\circ V}_{3})dT\},\\
di(\mathbf{V}_{4})G  &  =(\partial(\mathbf{H+V}_{3}\mathbf{\times D})/\partial
T+\nabla(\mathbf{H\circ V}_{3}))_{k}dT\symbol{94}dx^{k}+\\
&  \{curl(\mathbf{H+V}_{3}\mathbf{\times D})\}^{x}dy\symbol{94}dz\\
&  +\{curl(\mathbf{H+V}_{3}\mathbf{\times D})\}^{y}dz\symbol{94}dx\\
&  +\{curl(\mathbf{H+V}_{3}\mathbf{\times D})\}^{z}dx\symbol{94}dy\text{,
where}\\
(curl(\mathbf{H+V}_{3}\mathbf{\times D})  &  =\partial\mathbf{D}/\partial
T+\mathbf{J}_{3}+\mathbf{D\circ\lbrack}\nabla\mathbf{V}_{3}]-\mathbf{V}%
_{3}\circ\lbrack\nabla\mathbf{D]}+(\text{div}\mathbf{V}_{3})\mathbf{D-(}%
\text{div}\mathbf{D}\text{)}\mathbf{V}_{3},\\
i(\mathbf{V}_{4})dG  &  =(\mathbf{V}_{3}\times\mathbf{J}_{3})_{k}%
dT\symbol{94}dx^{k}\\
&  +(\mathbf{J}_{3}^{x}-\rho\mathbf{V}_{3}^{x})dy\symbol{94}dz+(\mathbf{J}%
_{3}^{y}-\rho\mathbf{V}_{3}^{y})dz\symbol{94}dx+(\mathbf{J}_{3}^{z}%
-\rho\mathbf{V}_{3}^{z})dx\symbol{94}dy.
\end{align}
If $L_{(\mathbf{V}_{4})}G$ is equal to zero, then there are 6 equations of
constraint:
\begin{align}
\partial(\mathbf{H+V}_{3}\mathbf{\times D})/\partial T+\nabla(\mathbf{H\circ
V}_{3})+(\mathbf{V}_{3}\times\mathbf{J}_{3})  &  =0,\\
\partial\mathbf{D}/\partial T+\mathbf{D\circ\lbrack}\nabla\mathbf{V}%
_{3}]-\mathbf{V}_{3}\circ\lbrack\nabla\mathbf{D]}+(\text{div}\mathbf{V}%
_{3})\mathbf{D}  &  =0.
\end{align}
It is apparent that the last equation represents a diffusion equation for the
components of the vector $\mathbf{D}$.

The next concept is to examine conformal invariance of the 2-form, $F$. \ The
equations are similar to those above:%

\begin{equation}
L_{(\mathbf{V}_{4})}G=\varkappa G.
\end{equation}
The analysis leads to the constraints:%

\begin{align}
\partial(\mathbf{H+V}_{3}\mathbf{\times D})/\partial T+\nabla(\mathbf{H\circ
V}_{3})+(\mathbf{V}_{3}\times\mathbf{J}_{3})-\varkappa\mathbf{H}  &  =0,\\
\partial\mathbf{D}/\partial T+\mathbf{D\circ\lbrack}\nabla\mathbf{V}%
_{3}]-\mathbf{V}_{3}\circ\lbrack\nabla\mathbf{D]}+(\text{div}\mathbf{V}%
_{3})\mathbf{D-}\varkappa\mathbf{D}  &  =0.
\end{align}
\newline Again, the conformality factor, $\varkappa$, can represent a
homogeneity condition, or a fractal self-similarity condition under the
evolutionary process. \ As before, such processes are modifications of a
diffusion equation.

It is important to note that these equations represent the process constraints
that preserve the 2-forms of Affine Torsion $\approx\left\vert G\right\rangle
$ conformally.

\subsection{Period Integrals and Topological Quantization.}

Although the main interest of this article is associated with the field
properties of the Cosmological Vacuum, a few words are appropriate about the
topological defect structures (that will be treated in more detail in another
article). \ Of specific interest are those topological structures represented
by closed, but not exact, differential forms. \ Such exterior differential
forms are homogeneous of degree zero expressions. \ Such closed structures can
lead to deRham period integrals \cite{rmkperiods}, whose values have rational
ratios, when the integration chain, z1, is also closed, and not equal to a
boundary. \ For example, the Flux Quantum of EM theory is given by the closed
integral of the (electrodynamic) 1-form of Action:%

\begin{align}
\text{Flux quantum}  &  =%
%TCIMACRO{\tint \limits_{z1}}%
%BeginExpansion
{\textstyle\int\limits_{z1}}
%EndExpansion
A=%
%TCIMACRO{\toint }%
%BeginExpansion
{\textstyle\oint}
%EndExpansion
A=n\ \hbar/e\\
\text{ \ In domains where }F  &  =dA\Rightarrow0
\end{align}
This period integral is NOT\ dependent upon the electromagnetic field
intensities, $F$. \ Stokes theorem does not apply if the integration chain is
not a boundary. \ All integrals of exact forms over boundaries would yield zero.

As another example, the Period integral of quantized charge is given by the expression,%

\begin{align}
\text{Charge quantum}  &  =%
%TCIMACRO{\tiint _{z2}}%
%BeginExpansion
{\textstyle\iint_{z2}}
%EndExpansion
G\text{ }=n\ (\hbar/e)^{2}\\
\text{ \ In domains where }J  &  =dG\Rightarrow0.
\end{align}
The integration is over a closed chain, z2, which is not a boundary. \ Again,
the quantized Period integral does not depend upon the charge current
3-form,\ $J$, and the expression is valid only in domains where $J\Rightarrow
0.$ \ These same properties of topological quantization are universal ideas
independent from the notation, or some topological refinement to a specific
types of physical systems. \ Note that the concept of the Charge quantum
depends upon the existence of $G$ which in turn depends upon the existence of
Affine Torsion of the Cartan Connection $[\mathbb{C}]$ based on $[\mathbf{B}]$.

Similar results apply to the 3-forms of Topological Torsion and Topological Spin:

\begin{align}
\text{Chirality (Helicity) quantum}  &  =%
%TCIMACRO{\tiiint \limits_{z3}}%
%BeginExpansion
{\textstyle\iiint\limits_{z3}}
%EndExpansion
A\symbol{94}F\text{ }=n\ (\hbar/e)^{2}\ \\
\text{ \ In domains where }d(A\symbol{94}F)  &  =0.
\end{align}

\begin{align}
\text{Spin quantum}  &  =%
%TCIMACRO{\tiiint \limits_{z3}}%
%BeginExpansion
{\textstyle\iiint\limits_{z3}}
%EndExpansion
A\symbol{94}G\text{ }=n\ (\hbar)\ \\
\text{ \ In domains where }d(A\symbol{94}G)  &  =0.
\end{align}

\subsection{The Source of Charge and Spin}

A number of years ago, it became apparent to me that the origin of charge was
to be associated with topological structures of space time\ \cite{rmkpoincare}%
. \ The concepts exploited in that study assumed the existence of an impair
2-form of field intensities $G.$ \ Based on the developments above, it can be stated:

\begin{remark}
The theory of a Cosmological Vacuum asserts that the existence of charge is
dependent upon the Affine Torsion of the Cartan connection $[\mathbb{C}]$.
\end{remark}

This result is based upon the\ formal correspondence between equations of
electromagnetic field excitations $\left\vert G\right\rangle $ and the "Affine
torsion" coefficients (not the Cartan torsion coefficients) deduced for the
Cartan Connection matrix of 1-forms $\left[  \mathbb{C}\right]  $.%

\begin{equation}
\text{Excitation 2-forms }\left\vert \mathfrak{T}\right\rangle =\left[
\mathbb{C}\right]  \symbol{94}\left\vert dy\right\rangle \text{ Affine Torsion
2-forms.}%
\end{equation}
\ The impair 2 forms that compose the elements of $\left\vert \mathfrak{T}%
\right\rangle $ formally define the field excitations in terms of the
coefficients of "Affine torsion". \ The closed period integrals of those
(closed but not exact) components of $\left\vert \mathfrak{T}\right\rangle ,$
which are homogeneous of degree 0, lead to the deRham integrals with rational,
quantized, ratios. \ Such excitation 2-forms $\mathfrak{T}$ do not exist if
the coefficients of Affine torsion are zero.

The topological features of the Cosmological Vacuum are determined by the
structural properties of the Basis Frames, $\left[  \mathbb{B}\right]  ,$ and
the derived Cartan Connection matrix of 1-forms, $\left[  \mathbb{C}\right]
.$

\begin{center}
\textbf{The Topological Structure of the Cosmological Vacuum}

\textbf{ in terms of }$\left[  \mathbb{B}\right]  $, $\left[  \mathbb{C}%
\right]  $, and $[\Gamma]$:

\textbf{Mass \symbol{126} }Non Zero Curvature $\left[  \Phi\right]  \,\ $based
on metric generated Christoffel Connection $[\Gamma]$. \ $\left[  \Phi\right]
=[\Gamma]\symbol{94}[\Gamma]+d[\Gamma]$. \ Note that the Cartan Curvature
based on $\left[  \mathbb{C}\right]  $ is zero

\textbf{Charge\ \symbol{126}}Non Zero Affine Torsion of $\left\vert
\mathfrak{T}\right\rangle $ based on the Cartan right Connection $[C]$.

$\qquad\left\vert \mathfrak{T}\right\rangle =\left[  \mathbb{C}\right]
\symbol{94}\left\vert dy\right\rangle $. \ The Cartan Torsion is Zero, but the
Affine Torsion need not be Zero.
\end{center}

Where the presence of mass is recognized in terms of the Riemannian curvature
of the quadratic congruences of a Cosmological Vacuum, the presence of charge
is recognized in terms of the coefficients of Affine Torsion of a Cosmological Vacuum.

\subsection{Quadratic Congruences and Metrics}

Starting from the existence of a Linear Basis Frame, it is remarkable that
symmetric properties of $\left[  \mathbb{B}\right]  $\ can be deduced in terms
of a quadratic congruence (see p. 36 in Turnbull and Aitken \cite{TurnAiken}).
The quadratic congruence is related to the concept of strain in elasticity
theory, and is quite different from the linear definition of matrix symmetries
in terms of the sum of a matrix $\left[  \mathbb{B}\right]  $ and its
transpose $\left[  \mathbb{B}\right]  ^{T}$. \ The algebraic quadratic
congruence will be used to define 0compatible symmetric (metric) qualities in
terms of the structure of the Basis Frames, $\left[  \mathbb{B}\right]  $:%
\begin{equation}
\left[  g\right]  =\left[  \mathbb{B}\right]  ^{T}\circ\left[  \eta\right]
\circ\left[  \mathbb{B}\right]  .
\end{equation}
The matrix $\left[  \eta\right]  $ is a (diagonal)\ Sylvestor signature matrix
whose elements are $\pm1.$ \ Recall that in projective geometry the congruence
transformation based upon $\left[  \mathbb{B}\right]  ^{T}$ defines a
correlation, where the similarity transformation based upon $\left[
\mathbb{B}\right]  ^{-1}$ defines a collineation. \ Note that the ubiquitous
choice of an orthonormal basis frame, where $\left[  \mathbb{B}\right]
^{T}=\left[  \mathbb{B}\right]  ^{-1},$ would place limits the topological
generality of the concept of a Cosmological Vacuum.

In a later section it will be demonstrated how the similarity invariants of
the Basis Frame find use in representing thermodynamic phase functions
appropriate to the Cosmological Vacuum.

\begin{remark}
Note that this \textit{quadratic} (multiplicative) symmetry property is not
the equivalent to the\ (additive) symmetry property defined by the
\textit{linear} sum of the matrix $\left[  \mathbb{B}\right]  $ and its
transpose. \ 
\end{remark}

However, from the Basis Frame, it is also possible to construct a topological
exterior differential system that defines a quadratic form of the law of
differential closure. \ It then follows that, for $d\left[  \eta\right]  =0,$
\begin{align}
d\left[  g\right]   &  =d\left[  \mathbb{B}\right]  ^{T}\circ\left[
\eta\right]  \circ\left[  \mathbb{B}\right]  +\left[  g\right]  =\left[
\mathbb{B}\right]  ^{T}\circ\left[  \eta\right]  \circ d\left[  \mathbb{B}%
\right]  ,\\
d\left[  g\right]   &  =[\widetilde{\mathbb{C}}_{r}]\circ\left[  g\right]
+\left[  g\right]  \circ\left[  \mathbb{C}_{r}\right]  .\text{ }%
\end{align}
The fact that the differential $d\left[  g\right]  $ is the sum of two 1-forms
is a topological property, known as the metricity condition:%

\begin{equation}
\text{\textbf{Metricity condition:} \ }d\left[  g\right]  -[\widetilde
{\mathbb{C}}_{r}]\circ\left[  g\right]  -\left[  g\right]  \circ\left[
\mathbb{C}_{r}\right]  \Rightarrow0.
\end{equation}
The thermodynamic evolution of the metric can be expressed in terms of
Cartan's magic formula:%

\begin{align}
L_{(\mathbf{V}_{4})}\left[  g\right]   &  =i(\mathbf{V}_{4})d\left[  g\right]
\text{ thermodynamic evolution of the metric}\\
&  =\{i(\mathbf{V}_{4})[\widetilde{\mathbb{C}}_{r}]\}\circ\left[  g\right]
+\left[  g\right]  \circ\{i(\mathbf{V}_{4})\left[  \mathbb{C}_{r}\right]  \}.
\end{align}
It is apparent that the thermodynamic evolution of the metric can depend upon
the path, $\mathbf{V}_{4},$ as well as the connection, $\left[  \mathbb{C}%
_{r}\right]  $. \ It is apparent that if $d[g]=0$, then the thermodynamic
evolution of the metric is invariant. \ 

If the metricity condition is satisfied, then the metric is a thermodynamic
evolutionary invariant. This constraint is not presumed in this article.
\ Note that his equation is an exterior differential system, and therefor
defines topological properties. \ 

Compute the Christoffel connection, and its matrix of 1-forms, $\left[
\Gamma\right]  ,$\ from the quadratic "metric" matrix $\left[  g\right]  ,$
using the Levi-Civita-Christoffel formulas.%
\begin{align}
\text{Coefficients}  &  :\text{Christoffel Connection}\\
\Gamma_{ac}^{b}(\xi^{c})  &  =g^{be}\{\partial g_{ce}/\partial\xi^{a}+\partial
g_{ea}/\partial\xi^{c}-\partial g_{ac}/\partial\xi^{e}\},\\
\text{ }\left[  \Gamma\right]   &  =\left[  \Gamma_{ac}^{b}dy^{c}\right]
\text{ \ as a matrix of 1-forms}%
\end{align}
The Christoffel Connection also satisfies the metricity condition,%

\begin{equation}
\text{\textbf{Metricity condition:} \ }d\left[  g\right]  -\left[
\Gamma\right]  \circ\left[  g\right]  -\left[  g\right]  \circ\left[
\Gamma\right]  \Rightarrow0.
\end{equation}

\subsubsection{Thermodynamic Killing Vectors}

The concept of a homogeneous or conformal mapping is given by the expression,%

\begin{equation}
L_{(\mathbf{V}_{4})}\left[  g\right]  =i(\mathbf{V}_{4})d\left[  g\right]
=\kappa\left[  g\right]  .
\end{equation}
Process direction fields that have this property are defined as Killing
vectors. \ If $\kappa=0$, then the metric is said to be a process invariant of
the flow generated by the Killing vector. \ If $\kappa\neq0$, then the mapping
is said to be conformal. \ Note that if $\kappa=1$, the process indicates that
the thermodynamic evolution of the metric is self-similar.

\subsection{The vector of zero forms (Internal Energy)}

Once again consider the Lie differential with respect to a direction field $V
$, operating on the formula for differential closure%
\begin{align}
L_{(V)}(\left[  \mathbb{B}(y)\right]  \circ\left\vert dy^{a}\right\rangle )\
&  =L_{(V)}(\left\vert A^{a}\right\rangle )=i(V)d\left\vert A^{a}\right\rangle
+d(i(V)\left\vert A^{a}\right\rangle )\\
&  =i(V)\left\vert F^{a}\right\rangle +d(i(V)\left\vert A^{a}\right\rangle
)=\\
&  =\left\vert W^{a}\right\rangle +d\left\vert h^{a}\right\rangle .
\end{align}
From Koszul's theorem, $\left\vert W^{a}\right\rangle =i(V)d\left\vert
A^{a}\right\rangle $ is a covariant differential based on some (abstract)
connection (for each $a).$ \ Hence, the difference between the Lie
differential and the Covariant differential is the exact term,
$d(i(V)\left\vert A^{a}\right\rangle ):$%
\begin{equation}
L_{(V)}(\left\vert A^{a}\right\rangle )-i(V)d\left\vert A^{a}\right\rangle
=d(i(V)\left\vert A^{a}\right\rangle )=d\left\vert h^{a}\right\rangle .
\end{equation}
This equation is another statement of Cohomology, another exterior
differential system, where the difference of two non-exact objects is an exact differential.

From the topological formulation of thermodynamics in terms of Cartan's magic
formula \cite{Marsden},
\begin{align}
\text{ Cartan's Magic Formula }L_{(\rho\mathbf{V}_{4})}A  &  =i(\rho
\mathbf{V}_{4})dA+d(i(\rho\mathbf{V}_{4})A)\\
\text{First Law }  &  :W+dU=Q,\\
\text{Inexact Heat 1-form\ \ }Q  &  =W+dU=L_{(\rho\mathbf{V}_{4})}A\\
\text{Inexact Work 1-form\ }W  &  =i(\rho\mathbf{V}_{4})dA,\\
\text{Internal Energy \ }U  &  =i(\rho\mathbf{V}_{4})A,
\end{align}
Now consider particular process paths (defined by the directional field
$\rho\mathbf{V}_{4})$, and deduce that in the direction of the process path
\begin{align}
\text{ }i(\rho\mathbf{V}_{4})W  &  =0,\text{ }\\
\text{Work }  &  :\text{ is transversal;}\\
i(\rho\mathbf{V}_{4})Q  &  =i(\rho\mathbf{V}_{4})dU\text{ }\neq0\\
\text{ Heat }  &  :\text{is not transversal;}\\
\text{but if \ }i(\rho\mathbf{V}_{4})Q  &  =0,\text{ }\\
\text{the process }  &  :\text{is adiabatic.}%
\end{align}

It is the non-adiabatic components of a thermodynamic process that indicate
that there is a change of internal energy and hence an inertial force in the
direction of a process. \ This implies that the non-adiabatic processes are
inertial effects, and could be related to changes in mass.

Now to paraphrase statements and ideas from Mason and Woodhouse, (see p. 49
\ \cite{MW}), and \cite{Atiyah} :

\begin{remark}
"Then there is a Higgs field $\phi_{V}$ associated with each conformal Killing
vector $V$ $\in\mathfrak{h}$, (the Lie algebra of H){\small \ }which measures
the difference between the Covariant derivative along $V$ and the Lie
derivative along $V$."
\end{remark}

The implication is that the concept of a Higgs field represents the difference
between a process that is NOT dependent upon the constraint of a gauge group
(the Lie differential), and a process that is\ restricted to a specific choice
of \ a connection defined by some gauge group, (the Covariant differential). \ 

It becomes apparent that:
\begin{align}
\left\vert W^{a}\right\rangle  &  =\text{Vector of Work 1-forms.
(transversal)}\\
\left\vert h^{a}\right\rangle  &  =\text{Higgs potential as vector of 0-forms
(Internal Energy)}\\
d\left\vert h^{a}\right\rangle  &  =\text{Higgs vector of 1-forms.}\\
i(V)d\left\vert h^{a}\right\rangle  &  =\text{vector of longitudinal inertial
accelerations (with mass)}\\
&  =\text{non adiabatic components of a process}%
\end{align}
The method of the "Cosmological Vacuum" and its sole assumption\ leads to
inertial properties and the Higgs field, all from a topological perspective
and without "quantum" fluctuations.\ 

\subsection{A Strong Equivalence Principle}

At this point, there has been no indication that the problem being
investigated has anything to do with the Gravitational Field. \ The gravity
issue is to be encoded into how the quadratic congruent symmetries of $\left[
\mathbb{B}\right]  ,$ and its topological group structures, are established.
\ In general, different choices for the group structure of the Basis Frame
will strongly influence the application to any particular physical system of
fields and particles. \ 

Without the Einstein Ansatz, it appears that the concept of a Cosmological
Vacuum can lead to a Strong Equivalence principle. \ Substitute $\left[
\Gamma\right]  +\left[  \mathbb{T}\right]  $ for $\left[  \mathbb{C}\right]  $
in the definition of the matrix of curvature 2-forms, and recall that for the
Cosmological Vacuum the Cartan matrix of curvature 2-forms, $[\Theta],$ is
zero.%
\begin{align}
\lbrack\Theta]  &  =\{d\left[  \mathbb{C}\right]  +\left[  \mathbb{C}\right]
\symbol{94}\left[  \mathbb{C}\right]  \}\Rightarrow0,\\
&  =\{d(\left[  \Gamma\right]  +\left[  \mathbb{T}\right]  )+(\left[
\Gamma\right]  +\left[  \mathbb{T}\right]  )\symbol{94}(\left[  \Gamma\right]
+\left[  \mathbb{T}\right]  )\}\\
&  =\{d[\Gamma]+[\Gamma]\symbol{94}[\Gamma]\}+\{\left[  \mathbb{T}\right]
\symbol{94}\left[  \Gamma\right]  +\left[  \Gamma\right]  \symbol{94}\left[
\mathbb{T}\right]  \}+\{d[\mathbb{T}]+[\mathbb{T}]\symbol{94}\mathbb{[T}]\},
\end{align}

Separate the matrices of 2-forms into the metric based (Christoffel) curvature
2-forms, defined as
\begin{equation}
\left[  \Phi_{\mathbf{\Gamma}}\right]  =\{d\left[  \Gamma\right]  +\left[
\Gamma\right]  \symbol{94}\left[  \Gamma\right]  \}=\left[
Field\ metric\ 2-forms\right]  ,
\end{equation}
and the remainder, defined as
\begin{align}
\left[  -\Sigma_{Inertial}\right]   &  =\left[  \Theta\right]  -\left[
\Phi_{\mathbf{\Gamma}}\right] \\
&  =\{\left[  \mathbb{T}\right]  \symbol{94}\left[  \Gamma\right]  +\left[
\Gamma\right]  \symbol{94}\left[  \mathbb{T}\right]  \}+\{d[\mathbb{T}%
]+[\mathbb{T}]\symbol{94}\mathbb{[T}]\}\\
&  =\{interaction\_2-forms\}\ \ +\ \{\left[  \Sigma_{\mathbb{T}}\right]  \}
\end{align}
$.$ The decomposition leads to the strong equivalence equation,%
\begin{align}
\text{\textbf{Principle of} }  &  \text{:\textbf{\ Strong Equivalence }}\\
\left[  \text{Metric Field curvature 2-forms}\right]   &  =\left[
\text{Inertial curvature 2-forms}\right]  ,\\
\left[  \Phi_{\mathbf{\Gamma}}\right]   &  =\left[  -\Sigma_{Inertial}\right]
\end{align}

The Cartan Connection is not necessarily equal to the metric connection, but
the above formula generates a vector of 2-forms that are equal, for any
decomposition. \ If the Basis Frame is chosen such that the Cartan Connection
is equal to the Christoffel connection, then the residue term vanishes. \ This
result implies that the vector of affine torsion 2-forms vanishes. \ In the
general case, the decomposition of the right Cartan connection can have
symmetric as well as anti-symmetric parts, but it is always the case that the
antisymmetric part generates the 2-forms of Affine Torsion, and therefor the
excitation 2-forms, $\left\vert G\right\rangle $. \ It is suggested that in
the choice of a symmetric basis frame such that the Cartan connection \ is
equal to the metric, with $[\mathbb{T}]=0$, corresponds to matter free space,
analogous to the Einstein metric based equations.

\section{Remarks}

This universal set ideas enumerated above startles me. \ There are only two assumptions:

\begin{enumerate}
\item \ The postulate that the domain of continuum fields is a vector space
defined by infinitesimal mappings.

\item \ The postulate that the field equations are generated by processes that
must satisfy the concepts of thermodynamic evolution ($\approx$ the First Law
of Thermodynamics as expressed in terms of continuous topological evolution
and Cartan's magic formula). \ \ 
\end{enumerate}

The rest of the concepts are derived, following the rules of the Cartan
exterior calculus. \ These results appear to be universal rules.
\ \ Particulate concepts also appear, from the same fundamental postulates of
a Cosmological Vacuum, in the form of topologically coherent defect structures
in the fields. \ Quantization occurs in a topological manner from the deRham
theorems as period integrals. \ The quantized topologically coherent
structures form the basis of macroscopic quantum states.\ 

Earlier, related, thoughts about the topological and differential geometric
ideas associated with the four forces appeared in \cite{rmksubmersive}. \ Not
only do the long range concepts (fields) of gravity and electromagnetism have
the a thermodynamic base, but so also do the short range concepts (fields) of
the nuclear and weak force. \ The short range forces are artifacts of
non-equilibrium thermodynamic systems with PTD\
%TCIMACRO{\TEXTsymbol{>} }%
%BeginExpansion
$>$
%EndExpansion
2, where the long range forces are artifacts of thermodynamic systems with
PTD
%TCIMACRO{\TEXTsymbol{<} }%
%BeginExpansion
$<$
%EndExpansion
3. \ \ The topological theory of a Cosmological Vacuum, as presented above,
re-enforces this earlier work. \ 

It is also remarkable that the methods described above also have applications
in engineering format, which can lead to new applications governed by the
non-equilibrium and perhaps chiral features of thermodynamic evolution. \ The
field theory of the Cosmological Vacuum appears to be applicable at all
scales. \ In electromagnetic language, the key design principle to produce
stable plasmas is to minimize $\mathbf{E\circ B}$.\ 

An important result is the recognition that the 2-forms of Affine torsion are
in effect the source of the topologically distinct Excitation 2-forms,
$\left\vert \mathfrak{T}\right\rangle $ and $\left\vert G\right\rangle $. \ It
follows that the topological foundation of both hydrodynamics and
electrodynamics can be put on the same footing. \ Yet $\left\vert
\mathfrak{T}\right\rangle $ the pair analogue of the impair $\left\vert
G\right\rangle $ does not appear in the classic literature of hydrodynamics.
\ Why? The simplest reason is the dogmatic adherence to symmetric metric
representations of stress and strain. \ Such methods eliminate the
possibilities of Affine Torsion in the classical theories. \ As another
example, the symmetric metric of GR\ theory utilizes Christoffel connections
which are free of Affine Torsion, again eliminating the important
thermodynamic concept of additive variables. \ However, Fluids and
Electromagnetism, based on a Cartan connection generated by infinitesimal
mappings, indeed can have the properties associated with Affine Torsion. \ In
fluids, the Affine Torsion is the source of unit mass (mole), and in
electromagnetism, Affine Torsion is the source of unit charge. \ However,
remember that the Affine Torsion 2-forms are necessary (but not sufficient)
for constructing the 3-forms of Topological Spin.

\section{Examples}

\subsection{ Example 1. \ \textbf{The Schwarzschild Metric embedded in a Basis
Frame,\ [}$\mathbb{B}$\textbf{], as a 10 parameter subgroup of an affine
group.}}

\subsubsection{The Metric - a Quadratic Congruent symmetry}

The algebra of a quadratic congruence can be used to deduce the metric
properties of a given Basis Frame. \ These deduced metric features may be used
to construct a "Christoffel" or metric compatible connection, different from
the Cartan Connection. \ The Christoffel connection constructed from the
quadratic congruence of the Basis Frame may or may not generate a "Riemannian"
curvature. \ By working backwards, this example will demonstrate how to
construct the Basis Frame, given a diagonal metric. \ The method is algebraic
and exceptionally simple for all diagonal metrics that represent a 3+1
division of space-time. \ The important result is that given a Basis Frame of
a given matrix group structure, a metric and a compatible Christoffel
Connection can be deduced. \ An important and unexpected result is that a
basis frame compatible with a metric as a quadratic congruent symmetry, can
generate a Cartan matrix that supports Affine Torsion.

\begin{remark}
All 3+1 metric structures are to be associated with the 10 parameter subgroup
of the 13 parameter Affine groups. \ 
\end{remark}

In this example, it will be demonstrated how the isotropic form of the
Schwarzschild metric can be incorporated into the Basis Frame for a Physical
Vacuum, $\left[  \mathbb{B}\right]  $. \ The technique is easily extendable
for diagonal metrics. \ However, the symmetry properties of the Cartan
Connection are not limited to metrics of the "gravitational" type. \ Once the
Schwarzschild metric is embedded in to the Basis Frame, then the universal
methods described above will be applied to the representative Basis Frame, and
each important result will be evaluated.

The isotropic Schwarzschild metric is a diagonal metric of the form,%

\begin{align}
(\delta s)^{2}  &  =-(1+m/2r)^{4}\{(dx)^{2}+(dy)^{2}+(dz)^{2}\}+\frac
{(2-m/r)^{2}}{(2+m/r)^{2}}(dt)^{2}\\
&  =-(\alpha)^{2}\{(dx)^{2}+(dy)^{2}+(dz)^{2}\}+(\beta)^{2}(dt)^{2}\\
\text{with }r  &  =\sqrt{(x)^{2}+(y)^{2}+(z)^{2}},
\end{align}
As Eddington \cite{EddingtonREL} points out, the isotropic form is palatable
with the idea that the speed of light is equivalent in any direction. \ That
is not true for the non-isotropic Schwarzschild metric, where transverse and
longitudinal null geodesics do not have the same speed. \ \ 

For the isotropic Schwarzschild example, the metric $\left[  g_{jk}\right]
$\ can be constructed from the triple matrix product:
\begin{align}
\left[  g_{jk}\right]   &  =[\widetilde{f}]\circ\left[  \eta\right]
\circ\left[  f\right]  ,\\
\text{where }f  &  =\left[
\begin{array}
[c]{cccc}%
\alpha & 0 & 0 & 0\\
0 & \alpha & 0 & 0\\
0 & 0 & \alpha & 0\\
0 & 0 & 0 & \beta
\end{array}
\right]  ,\\
\text{and \ \ \ }\alpha &  =(1+m/2r)^{2}\ =(\gamma/2r)^{2},\ \ \ \ \beta
=\frac{(2-m/r)}{(2+m/r)}\ =\delta/\gamma,\\
\text{and \ \ \ }\eta &  =\left[
\begin{array}
[c]{cccc}%
-1 & 0 & 0 & 0\\
0 & -1 & 0 & 0\\
0 & 0 & -1 & 0\\
0 & 0 & 0 & 1
\end{array}
\right]  .
\end{align}
At first glance it would appear that the Schwarzschild metric forms a
quadratic form constructed from the congruence of a 4 parameter (diagonal)
matrix. \ It is not obvious that this may be a special case of a 10 parameter
group with a fixed point. \ In order to admit the 10 parameter Poincare group
(which is related to the Lorentz group), a map from spherical 3+1 space to
Cartesian 3+1 space will be perturbed by the matrix $[f].$

\subsubsection{The Diffeomorphic Jacobian Basis Frame}

At first, consider the diffeomorphic map $\phi^{k}$ from spherical\ 3+1 space
to Cartesian 3+1 coordinates:
\begin{align}
\{y^{a}\}  &  =\{r,\theta,\varphi,\tau\}\Rightarrow\{x^{k}\}=\{x,y,z,t]\\
\phi^{k}  &  :[r\sin(\theta)\cos(\varphi),r\sin(\theta)\sin(\varphi
),rcos(\theta),\tau]\Rightarrow\lbrack x,y,z,t]\\
\{dy^{a}\}  &  =\{dr,d\theta,d\phi.d\tau\}.
\end{align}
The Jacobian of the diffeomorphic map $\phi^{k}$ can be utilized as an
integrable Basis Frame matrix\ $\left[  \mathbb{B}\right]  $ which is an
element of the 10 parameter F-Affine group (The Affine subgroup with a fixed point):%

\begin{equation}
\left[  \mathbb{B}\right]  =\left[
\begin{array}
[c]{cccc}%
\sin(\theta)\cos(\varphi) & r\cos(\theta)\cos(\varphi) & -r\sin(\theta
)\sin(\varphi) & 0\\
\sin(\theta)\sin(\varphi) & r\cos(\theta)\sin(\varphi) & r\sin(\theta
)\cos(\varphi) & 0\\
\cos(\theta) & -r\sin(\theta) & 0 & 0\\
0 & 0 & 0 & 1
\end{array}
\right]  .
\end{equation}
\bigskip The infinitesimal mapping formula based on $\left[  \mathbb{B}%
\right]  $ yields%

\begin{equation}
\left[  \mathbb{B}_{a}^{k}(y)\right]  \circ\left\vert
\begin{array}
[c]{c}%
dr\\
d\theta\\
d\varphi\\
dt
\end{array}
\right\rangle \ \Rightarrow\ \left\vert \sigma^{k}\right\rangle =\left\vert
\begin{array}
[c]{c}%
dx\\
dy\\
dz\\
dt
\end{array}
\right\rangle ,
\end{equation}
such that all of the 1-forms $\left\vert \sigma^{k}\right\rangle $ are exact
differentials, and there are no non-zero field intensities, $\left\vert
d\sigma^{k}\right\rangle .$

\subsubsection{The Perturbed Basis Frame with a Congruent Symmetry}

\begin{theorem}
\ \ \textit{The effects of a diagonal metric }$\left[  g_{jk}\right]
$\textit{\ can be absorbed into a re-definition of the Frame matrix:}%
\begin{equation}
\lbrack\widehat{\mathbb{B}}]=\left[  f\right]  \circ\lbrack\mathbb{B}].\
\end{equation}

\end{theorem}

The integrable Jacobian Basis Frame matrix given above will be perturbed by
multiplication on the left by the diagonal matrix, $\left[  f\right]  .$ \ The
perturbed Basis Frame becomes%

\begin{align}
\lbrack\widehat{\mathbb{B}}]  &  =\left[  f\right]  \circ\lbrack
\mathbb{B}]\text{ \ the Schwarzschild Cartan Basis Frame.}\\
&  =\left[
\begin{array}
[c]{cccc}%
\sin(\theta)\cos(\varphi)\gamma^{2}/4r^{2} & \cos(\theta)\cos(\varphi
)\gamma^{2}/4r & -\sin(\theta)\sin(\varphi)\gamma^{2}/4r & 0\\
\sin(\theta)\sin(\varphi)\gamma^{2}/4r^{2} & \cos(\theta)\sin(\varphi
)\gamma^{2}/4r & \sin(\theta)\cos(\varphi)\gamma^{2}/4r & 0\\
\cos(\theta)\gamma^{2}/4r^{2} & -\sin(\theta)\gamma^{2}/4r & 0 & 0\\
0 & 0 & 0 & \delta/\gamma
\end{array}
\right]  .\ \\
\gamma &  =(2r+m),\ \ \ \ \delta=(2r-m)
\end{align}
Use of the congruent pullback formula based on the perturbed Basis Frame,
$[\widehat{\mathbb{B}}]$, yields,%

\begin{align}
\left[  g_{jk}\right]   &  =[\widehat{\mathbb{B}}_{transpose}]\circ\eta
\circ\lbrack\widehat{\mathbb{B}}],\\
\left[  g_{jk}\right]   &  =\left[
\begin{array}
[c]{cccc}%
-(\gamma^{2}/4r^{2})^{2} & 0 & 0 & 0\\
0 & -(\gamma^{2}/4r)^{2} & 0 & 0\\
0 & 0 & -(\gamma^{2}/4r)^{2}\sin^{2}(\theta) & 0\\
0 & 0 & 0 & +(\delta/\gamma)^{2}%
\end{array}
\right]  ,\\
\gamma &  =(2r+m),\ \ \ \ \delta=(2r-m)
\end{align}
which agrees with formula given above for the isotropic Schwarzschild metric
in spherical coordinates. \ It actually includes a more general idea, for the
coefficients, $\alpha,$and $\beta,$ can be dependent upon both $r$ and $\tau$. \ 

The infinitesimal mapping formula based on $[\widehat{\mathbb{B}}]$ yields%

\begin{equation}
\left[  \mathbb{B}_{a}^{k}(y)\right]  \circ\left\vert
\begin{array}
[c]{c}%
dr\\
d\theta\\
d\varphi\\
dt
\end{array}
\right\rangle \ \Rightarrow\ \left\vert \sigma^{k}\right\rangle =\left\vert
\begin{array}
[c]{c}%
(\gamma^{2}/4r^{2})dx\\
(\gamma^{2}/4r^{2})dy\\
(\gamma^{2}/4r^{2})dz\\
(\delta/\gamma)dt
\end{array}
\right\rangle ,
\end{equation}
such that all of the 1-forms $\left\vert \sigma^{k}\right\rangle $ are NOT
exact differentials. \ The 2-forms of field intensities, $\left\vert
d\sigma^{k}\right\rangle $ are not zero.

\subsubsection{The Schwarzschild-Cartan connection.}

The Schwarzschild-Cartan (right) Connection $[\widehat{\mathbb{C}}],$ as a
matrix of 1-forms relative to the perturbed Basis Frame $[\widehat{\mathbb{B}%
}],$ becomes%

\begin{align}
\lbrack\widehat{\mathbb{C}}]  &  =[\widehat{\mathbb{B}^{-1}}]\circ
d[\widehat{\mathbb{B}}],\\
\lbrack\widehat{\mathbb{C}}]  &  =\left[
\begin{array}
[c]{cccc}%
-2mdr/r\gamma & -rd\theta & \sin^{2}(\theta)rd\phi & 0\\
d\theta/r & \delta dr/\gamma & -\cos(\theta)\sin(\theta)d\phi & 0\\
d\phi/r & \cot(\theta)d\phi & \cot(\theta)d\theta+\delta dr/\gamma & 0\\
0 & 0 & 0 & 4mdr/(\gamma\delta)
\end{array}
\right]  .\\
\gamma &  =(2r+m),\ \ \ \ \delta=(2r-m)
\end{align}

\begin{center}
\textbf{Perturbed Cartan Connection\bigskip\ from Maple}
\end{center}

It is apparent that the Cartan Connection matrix of 1-forms is again a member
of the 10 parameter matrix group, a subgroup of the 13 parameter affine matrix group.

\subsubsection{Vectors of Closure 2-forms}

Surprisingly, for the perturbed Basis Frame $[\widehat{\mathbb{B}}]$\ which
contains a the square root of a congruent metric field of a massive object,
the vector of "excitation" torsion 2-forms, based on the 10 parameter affine
subgroup with a fixed point, is not zero, and can be evaluated as:%

\begin{align}
\widehat{\left\vert \mathfrak{T}\right\rangle }  &  =[\widehat{\mathbb{C}%
}]\symbol{94}\left\vert dy^{a}\right\rangle \text{ \ }\\
\text{Torsion 2-forms }  &  :\text{\ of the Affine subgroup with a fixed
point}\\
\widehat{\left\vert \mathfrak{T}\right\rangle }  &  =\left\vert
\begin{array}
[c]{c}%
0\\
(2m/r\gamma)(d\theta\symbol{94}dr)\\
(2m/r\gamma)(d\phi\symbol{94}dr)\\
(4m/\gamma\delta)(dr\symbol{94}d\tau)
\end{array}
\right\rangle \text{ "Schwarzschild Excitations" }\\
\gamma &  =(2r+m),\ \ \ \ \delta=(2r-m)
\end{align}
The unexpected result is that the isotropic Schwarzschild metric admits
coefficients of "Affine Torsion" relative to the Cartan Connection matrix,
$[\widehat{\mathbb{C}}]$.

Similarly the vector of 2-form of field intensities $\widehat{\left\vert
F\right\rangle }$ can be evaluated in terms of the perturbed Basis Frame as:%
\begin{align}
\widehat{\left\vert d\sigma\right\rangle }  &  =d([\widehat{\mathbb{B}}%
]\circ\left\vert dy^{a}\right\rangle )\text{ \ \ \ \ \ "Schwarzschild
Intensities" }\\
&  :\text{Intensity 2-forms of the Affine subgroup with a fixed point}\\
\widehat{\left\vert d\sigma\right\rangle }  &  =\left\vert
\begin{array}
[c]{c}%
+(m\gamma/2r^{2})\{(\sin(\phi)\cos(\theta)d\theta\symbol{94}dr)-(\sin
(\theta)\cos(\phi)d\phi\symbol{94}dr)\\
+(m\gamma/2r^{2})\{(\sin(\phi)\cos(\theta)d\theta\symbol{94}dr)+(\sin
(\theta)\cos(\phi)d\phi\symbol{94}dr)\\
-(m\gamma/2r^{2})(\sin(\theta)d\theta\symbol{94}dr)\\
(4m/\gamma^{2})(dr\symbol{94}d\tau)
\end{array}
\right\rangle .\\
\gamma &  =(2r+m),\ \ \ \ \delta=(2r-m)
\end{align}

The constitutive map relating the field intensities and the field excitations,%

\begin{equation}
\widehat{\left\vert \mathfrak{T}\right\rangle }=[\widehat{\mathbb{B}}%
]^{-1}\circ\widehat{\left\vert d\sigma\right\rangle },
\end{equation}
is determined by the inverse of the perturbed Basis Frame, $[\widehat
{\mathbb{B}}]^{-1}$ :

\begin{center}%
\begin{align}
&  \text{\textbf{Schwarzschild Constitutive map from Maple}}\\
\lbrack\widehat{\mathbb{B}}]^{-1}  &  =4r/\gamma^{2}\left[
\begin{array}
[c]{cccc}%
r\sin(\theta)\cos(\phi) & r\sin(\theta)\sin(\phi) & \cos(\theta) & 0\\
\cos(\theta)\cos(\phi) & \cos(\theta)\sin(\phi) & -\sin(\theta) & 0\\
-\frac{\sin(\phi)}{\sin(\theta)} & \frac{\cos(\phi)}{\sin(\theta)} & 0 & 0\\
0 & 0 & 0 & \gamma^{3}/4r\delta
\end{array}
\right] \\
\gamma &  =(2r+m),\ \ \ \ \delta=(2r-m)
\end{align}

\end{center}

\subsubsection{Vectors of 3-forms}

\ The exterior derivative of the vector of excitations (Affine Torsion
2-forms) is zero, hence there are no current 3-forms of excitations,
$\left\vert J\right\rangle $, for the perturbed Basis Frame that encodes the
Schwarzschild metric as a Congruent symmetry:%

\begin{equation}
\text{Charge Current 3-form \ \ }\left\vert J\right\rangle =d\left\vert
\mathfrak{T}\right\rangle =0.
\end{equation}

The Topological Torsion 3-form for this isotropic Schwarzschild example
vanishes: $H=\left\vert A\symbol{94}F\right\rangle $\ $\Rightarrow0.\ $The
implication is that the system is of Pfaff dimension 2 (and therefor is an
equilibrium thermodynamic system).

Both Poincare 4-forms vanish, but the Topological Spin 3-form is NOT\ zero.
\ The individual components are:%

\begin{equation}
\widehat{\left\vert \sigma\symbol{94}\mathfrak{T}\right\rangle }=\left\vert
\begin{array}
[c]{c}%
0\\
-m(2r+m)\cos(\phi)\sin(\theta)(d\theta\symbol{94}d\phi\symbol{94}%
dr)/(2r^{2})\\
-m(2r+m)sin(\theta)(d\theta\symbol{94}d\phi\symbol{94}dr)/(2r^{2})\\
0
\end{array}
\right\rangle \text{ "Schwarzschild Spin components" }%
\end{equation}
The inner product of the components of potential 1-forms, $\left\langle
A\right\vert $ and the Excitation 2-forms yields:%

\begin{align}
&  \text{Total Topological Spin 3-form \ \ \ }\\
\text{\ }\mathfrak{S}  &  =\left\langle \sigma\right\vert \symbol{94}%
\left\vert \mathfrak{T}\right\rangle \ \ \ \ \ dA=0.\\
&  =(-m\gamma\sin(\theta)/2r^{2})\{\cos(\phi)+1)(d\theta\symbol{94}%
d\phi\symbol{94}dr)\}.
\end{align}
The "Topological Spin" 3-form depends upon the "mass" coefficient, $m$, and
the 2-forms of Affine torsion.

\subsubsection{The three Connection matrices}

The three matrices of Connection 1-forms are presented below for each
(perturbed) connection, $[\Gamma],~[\mathbb{C}],~[\mathbb{T]}$%

\begin{align}
\lbrack\widehat{\mathbf{\Gamma}}]  &  =\left[
\begin{array}
[c]{cccc}%
-2mdr/r\gamma & -\delta rd\theta/\gamma & \delta\sin^{2}(\theta)rd\phi/\gamma
& 64\delta md\tau/\gamma^{7}\\
\delta d\theta/r\gamma & \delta dr/r\gamma & -\cos(\theta)\sin(\theta)d\phi &
0\\
\delta d\phi/r\gamma & \cot(\theta)d\phi & \cot(\theta)d\theta+\delta
dr/\gamma & 0\\
4md\tau/(\gamma\delta) & 0 & 0 & 4mdr/(\gamma\delta)
\end{array}
\right]  .\\
\lbrack\widehat{\mathbb{C}}]  &  =\left[
\begin{array}
[c]{cccc}%
-2mdr/r\gamma & -rd\theta & \sin^{2}(\theta)rd\phi & 0\\
d\theta/r & \delta dr/\gamma & -\cos(\theta)\sin(\theta)d\phi & 0\\
d\phi/r & \cot(\theta)d\phi & \cot(\theta)d\theta+\delta dr/\gamma & 0\\
0 & 0 & 0 & 4mdr/(\gamma\delta)
\end{array}
\right]  .\\
\lbrack\widehat{\mathbb{T}}]  &  =\left[
\begin{array}
[c]{cccc}%
0 & -2mrd\theta/\gamma & 2m\sin^{2}(\theta)rd\phi/\gamma & -64mr^{4}\delta
d\tau/\gamma^{7}\\
2md\theta/r\gamma & 0 & 0 & 0\\
4md\phi/r\gamma & 0 & 0 & 0\\
-4md\tau/\delta\gamma & 0 & 0 & 0
\end{array}
\right]  .\\
\gamma &  =(2r+m),\ \ \ \ \delta=(2r-m)
\end{align}

\begin{center}
\textbf{Schwarzschild Perturbed Connections}
\end{center}

The matrix of (metric) curvature 2-forms, $\left[  \Phi_{\Gamma}\right]  ,$
based on the formula%

\begin{equation}
\left[  \Phi_{\Gamma}\right]  =d[\Gamma]+[\Gamma]\symbol{94}[\Gamma],
\end{equation}

\begin{center}
is computed to be:%
\begin{align}
&
\text{\textbf{Curvature\ 2-forms\ for\ the\ Schwarzschild\ Christoffel\ Connection}%
}\\
\lbrack\Phi_{\Gamma}]  &  =4m/\gamma^{2}\left[
\begin{array}
[c]{cccc}%
0 & -rdr\symbol{94}d\theta & r\sin^{2}(\theta)dr\symbol{94}d\phi &
-32r^{4}dr\symbol{94}d\tau/\gamma^{6}\\
2mdr\symbol{94}d\theta/r\gamma & 0 & -2\sin^{2}(\theta)d\theta\symbol{94}d\phi
& 16\delta^{2}r^{3}d\theta\symbol{94}d\tau/\gamma^{6}\\
4mdr\symbol{94}d\phi/r\gamma & -2rd\theta\symbol{94}d\phi & 0 & 16\delta
^{2}r^{3}d\phi\symbol{94}d\tau/\gamma^{6}\\
-4mdr\symbol{94}d\tau/\delta\gamma & 2rd\theta\symbol{94}d\tau & -2\sin
^{2}(\theta)d\phi\symbol{94}d\tau & 0
\end{array}
\right]
\end{align}

\end{center}

By the Strong Equivalence Principle,%

\begin{align}
&  :\{d[\Gamma]+[\Gamma]\symbol{94}[\Gamma]\}\ \ \ \ +\{\left[  \mathbb{T}%
\right]  \symbol{94}\left[  \Gamma\right]  +\left[  \Gamma\right]
\symbol{94}\left[  \mathbb{T}\right]  \}\ \ \ +\{d[\mathbb{T}]+[\mathbb{T}%
]\symbol{94}\mathbb{[T}]\}\\
&  =\ \ \ \ \ \{\left[  \mathbf{\Phi}_{\Gamma}\right]
\}\ \ \ \ \ \ +\{\left[  Interaction\ 2-forms\right]  \}+\ \ \ \{\left[
\mathbf{\Phi}_{\mathbf{T}}\right]  \}\Rightarrow0,
\end{align}
which can be checked using Maple.

\subsubsection{Summary Remarks}

The idea that has been exploited is that the arbitrary Basis Frame (a linear
form), without metric, can be perturbed algebraically to produce a new Basis
Frame that absorbs the properties of a quadratic congruent metric system.
\ This result establishes a constructive existence proof that compatible
metric features of a Physical Vacuum can be derived from the structural format
of the Basis Frame. \ The Basis Frame is the starting point and the congruent
metric properties are deduced.

For the Schwarzschild example, another remarkable feature is that the 1-forms
$\left\vert \sigma^{k}\right\rangle $ constructed according to the formula
\begin{equation}
\lbrack\widehat{\mathbb{B}}]\circ\left\vert dy^{a}\right\rangle \Rightarrow
\left\vert \sigma^{k}\right\rangle \approx\left\vert A^{k}\right\rangle ,
\end{equation}
are all integrable (as the Topological Torsion term is Zero), but the
coefficients of affine torsion are not zero. \ The symbol $\left\vert
dy^{a}\right\rangle $ stands for the set $[dr,d\theta,d\varphi,d\tau]$
(transposed into a column vector), and $[\widehat{\mathbb{B}}]$ is the
"perturbed" Basis Frame which contains the Schwarzschild metric as a congruent
symmetry. \ The integrability condition means that there exist integrating
factors $\lambda^{(k)}$ for each $\sigma^{k}$ such that a new Basis Frame can
be constructed from\ $[\widehat{\mathbb{B}}]$ algebraically. \ Relative to
this new Basis Frame, the vector of torsion 2-forms is zero, $\left\vert
d\sigma^{k}\right\rangle =\left\vert dA^{k}\right\rangle =\left\vert
F^{k}\right\rangle =0$! \ The "Coriolis" acceleration which is related to the
2-form of torsion 2-forms $\left\vert d\sigma^{k}\right\rangle $ can be
eliminated algebraically.! \ Although this result is possible algebraically,
it is not possible diffeomorphically. \ 

\ Of course, this algebraic reduction is impossible if any of the 1-forms,
$\sigma^{k}$, is of Pfaff dimension 3 or more. \ The Basis Frame then admits
Topological Torsion, which is irreducible. \ 

\subsection{\textbf{Example 2: \ [}$B$\textbf{] as a 13 parameter matrix
group}}

\subsubsection{The Intransitive "Wave -Affine" Basis frame}

The next set of examples considers the structure of those 4 x 4 Basis Frames
that admit a 13 parameter group in \ 4 geometrical dimensions of space-time.
\ There are 3 interesting types of 13 parameter group structures. \ This first
example utilizes the canonical form of the 13 parameter "Wave Affine" Basis
Frame. \ These Basis Frames will have zeros for the first 3 elements of the
right-most column. \ Wave Affine Basis Frames exhibit closure relative to
matrix multiplication. \ All products of Wave Affine Basis Frames have 3 zeros
on the right column. \ From a projective point of view, these matrices are not
elements of a transitive group. \ They are intransitive and have fixed points.
\ The true affine group in 4 dimensions is a transitive group (without fixed
points)\ and is discussed in \cite{rmkpv}.

For simplicity in display, the 9 parameter space-space portions of the Basis
Frame will be assumed to be the $3\times3$ Identity matrix, essentially
ignoring spatial deformations and spatially extended rigid body motions. \ In
the language of projective geometry, this intransitive system has a fixed
point. \ The first 3 elements in the bottom row can be identified (formally to
within a constant factor) with the components of a vector potential in
electromagnetic theory. The 4th (space-time) column will have three zeros, and
the $\mathbb{B}_{4}^{4}$ component will be described in terms of a function
$\phi(x,y,z,t)$. \ For convenience this "electrodynamic notation" for the
field intensities will be utilized.%

\begin{equation}
\text{ }\left[  \mathbb{B}_{wave\_affine}\right]  =\left[
\begin{array}
[c]{cccc}%
1 & 0 & 0 & 0\\
0 & 1 & 0 & 0\\
0 & 0 & 1 & 0\\
A_{x} & A_{y} & A_{z} & -\phi
\end{array}
\right]  .
\end{equation}

\subsubsection{The Field Potential 1-forms}

The projected 1-forms of potentials become%

\begin{equation}
\lbrack\mathbb{B}_{wave\_affine}]\circ\left\vert dy^{a}\right\rangle
=\left\vert A^{k}\right\rangle
\end{equation}%
\begin{align}
\left[
\begin{array}
[c]{cccc}%
1 & 0 & 0 & 0\\
0 & 1 & 0 & 0\\
0 & 0 & 1 & 0\\
A_{x} & A_{y} & A_{z} & -\phi
\end{array}
\right]  \circ\left\vert
\begin{array}
[c]{c}%
dx\\
dy\\
dz\\
dt
\end{array}
\right\rangle  &  \Rightarrow\left\vert
\begin{array}
[c]{c}%
dx\\
dy\\
dz\\
Action
\end{array}
\right\rangle =\left\vert A^{k}\right\rangle ,\\
Action  &  =A_{x}dx+A_{y}dy+A_{z}dz-\phi dt
\end{align}

\subsubsection{The Field Intensity 2-forms}

The exterior derivative of the vector of 1-forms, $\left\vert A^{k}%
\right\rangle $, produces the vector of 2-forms representing the field
intensities $\left\vert F^{k}\right\rangle $ of electromagnetic theory,
\begin{equation}
\text{\textbf{Intensity 2-forms:} \ }d\left\vert A^{k}\right\rangle
=\left\vert F^{k}\right\rangle =\left\vert
\begin{array}
[c]{c}%
dx\\
dy\\
dz\\
d(Action)
\end{array}
\right\rangle .
\end{equation}
Note that the Action 1-form produced by the wave affine Basis Frame is
precisely the format of the 1-form of Action used to construct the
Electromagnetic field intensities in classical EM theory.

\subsubsection{The Cartan Right Connection of 1-forms}

The Cartan right Connection matrix of 1-forms based upon the Basis Frame,
$\left[  \mathbb{B}_{wave\_affine}\right]  ,$ given above is given by the expression%

\begin{equation}
\left[  \mathbb{C}_{wave\_affine\_right}\right]  =\left[
\begin{array}
[c]{cccc}%
0 & 0 & 0 & 0\\
0 & 0 & 0 & 0\\
0 & 0 & 0 & 0\\
-d(A_{x})/\phi & -d(A_{y})/\phi & -d(A_{z})/\phi & d(\ln\phi)
\end{array}
\right]  .
\end{equation}

\subsubsection{The Field Excitation 2-forms (Affine Torsion)}

The general theory permits the vector of excitation 2-forms, $\left\vert
G\right\rangle $, to be evaluated as:%
\begin{align}
\left[  \mathbb{C}_{wave\_affine\_right}\right]  \symbol{94}\left\vert
dy^{m}\right\rangle  &  \simeq\left\vert G\right\rangle \text{
\ \ \ \ \ \ \ \textbf{Affine Torsion}}\\
\text{Excitation\textbf{ 2-forms} }  &  \left\vert G\right\rangle =\left\vert
\begin{array}
[c]{c}%
0\\
0\\
0\\
-(F)/\phi
\end{array}
\right\rangle =\left\vert
\begin{array}
[c]{c}%
0\\
0\\
0\\
-d(Action)/\phi
\end{array}
\right\rangle \\
&  \neq\left\vert \Sigma_{Car\tan\_Torsion}\right\rangle
\end{align}
The coefficients of the vector of excitation 2-forms are precisely those
ascribed to the coefficients of "Affine Torsion", even though the Basis Frame
used as the example is not a member of the transitive affine group.

\subsubsection{The Excitation Current 3-form}

The exterior derivative of the vector of excitation two forms $\left\vert
G\right\rangle $ produces the vector of 3-form currents $\left\vert
J\right\rangle $. \ The result is%

\begin{equation}
\left\vert J\right\rangle =\left\vert
\begin{array}
[c]{c}%
0\\
0\\
0\\
d(-F)/\phi
\end{array}
\right\rangle =\left\vert
\begin{array}
[c]{c}%
0\\
0\\
0\\
+d\phi\symbol{94}F/\phi^{2}%
\end{array}
\right\rangle
\end{equation}
Note that the charge-current 3-form is non-zero, but closed, producing the
expected Maxwell-Ampere charge conservation law:%

\begin{equation}
d\left\vert J\right\rangle =\left\vert 0\right\rangle
\end{equation}

The \textbf{wave} \textbf{affine group} of Basis Frames supports a vector of
"Affine Torsion" 2-forms that are abstractly related to excitation 2-forms
$\left\vert G\right\rangle $ representing the fields (the $\mathbf{D}$ and
$\mathbf{H}$ fields) generated by the sources in classical EM theory. \ The
charge-current 3-form need not vanish. \ 

\subsubsection{Topological Torsion and Topological Spin 3-forms}

The 3-forms of Topological Spin and Topological Torsion are proportional to
one another for the simple example, and are not zero if the Pfaff Topological
Dimension of the 1-form of Action is 3 or more. \ Such systems are not in
thermodynamic equilibrium.%
\begin{align*}
\text{Topological Torsion}  &  \text{: \ }A\symbol{94}F\\
\text{Topological Spin}  &  \text{: \ }A\symbol{94}G=-A\symbol{94}F/\phi
\end{align*}
If the Poincare invariants are to vanish it is necessary that the 4-form of
Topological Parity is zero, $F\symbol{94}F\Rightarrow0.$

\subsubsection{A time independent simplification}

For the simple time independent case, where, $Ax=0,\ Ay=0,\ Az(x,y,z)\neq
0,\ \phi(x,y,z)\neq0$, the vector of affine torsion 2-forms (the excitations) becomes:%

\begin{equation}
\text{Affine Torsion 2-forms \ \ }\left\vert G\right\rangle =\left\vert
\begin{array}
[c]{c}%
0\\
0\\
0\\
(-dA_{z}\symbol{94}dz+d\phi\symbol{94}dt)/\phi
\end{array}
\right\rangle ,
\end{equation}
and the 3-form of current induced becomes%

\begin{equation}
\text{Conserved Current 3-forms \ \ }\left\vert J\right\rangle =\left\vert
dG\right\rangle =\left\vert
\begin{array}
[c]{c}%
0\\
0\\
0\\
(dA_{z}\symbol{94}dz\symbol{94}d\phi)/\phi^{2}%
\end{array}
\right\rangle ,
\end{equation}
The vector of field intensity 2-forms becomes:%
\begin{equation}
\text{Field Intensity 2-forms \ }\left\vert F\right\rangle =\left\vert
\begin{array}
[c]{c}%
0\\
0\\
0\\
(-dA_{z}\symbol{94}dz+d\phi\symbol{94}dt)/\phi
\end{array}
\right\rangle ,
\end{equation}
The vector Topological Torsion 3-forms, becomes:%
\begin{equation}
\left\vert A\symbol{94}F\right\rangle =\left\vert
\begin{array}
[c]{c}%
0\\
0\\
0\\
(dA_{z}+A_{z}d\phi/\phi)\symbol{94}dz\symbol{94}dt)
\end{array}
\right\rangle ,
\end{equation}
The vector Topological Torsion 4-forms, becomes:
\begin{equation}
\text{Topological Torsion 4-forms \ \ \ }\left\vert F\symbol{94}F\right\rangle
=\left\vert
\begin{array}
[c]{c}%
0\\
0\\
0\\
(dA_{z}\symbol{94}d\phi/\phi)\symbol{94}dz\symbol{94}dt)
\end{array}
\right\rangle ,
\end{equation}

The second Poincare 4-form (the bulk viscosity "expansion" coefficient)
becomes proportional to the first Poincare 4-form,
\begin{equation}
d(A\symbol{94}F)=-2(\mathbf{E\circ B})\Omega_{4}=\phi\cdot d(A\symbol{94}G).
\end{equation}
Hence neither the Topological Spin, $A\symbol{94}G$, or the Topological
Torsion, $A\symbol{94}F$, define process direction fields that are leave the
Topological Torsion or the the Topological Spin as evolutionary invariants.
\ As the Topological Torsion 4-form is not zero, the system is of Pfaff
Topological Dimension 4, and is a non-equilibrium thermodynamic system.

Recall, that all of the analysis above is equally applicable to fluids which
admit Affine Torsion.

\textbf{Part II will consist of\ a collection of examples in terms of Maple
programs formatted as pdf files.}


\begin{thebibliography}{99}
\bibitem {Atiyah}Atiyah, M. : Collected Works by Michael Francis Atiyah -
Mathematics - 1988 - 720 pages Page 583 - "For this we just have to recall
that, intrinsically, the Higgs field is the difference between the Lie
derivative and the covariant derivative along the orbit\ldots"
\end{thebibliography}
\end{document}